\documentclass[fleqn,usenatbib]{mnras}

\usepackage{newtxtext,newtxmath}

\usepackage{xcolor,xspace}

\usepackage[T1]{fontenc}

\DeclareRobustCommand{\VAN}[3]{#2}
\let\VANthebibliography\thebibliography
\def\thebibliography{\DeclareRobustCommand{\VAN}[3]{##3}\VANthebibliography}

\usepackage{graphicx}	
\usepackage{amsmath}

\defcitealias{2011ApJ...728...88G}{GK11}

\newcommand{\tcm}{21$\,$cm \xspace~}  

\title[Studying galaxy cluster evolution with CHIME]{Forecasting Galaxy Cluster \ion{H}{i} Mass Recovery with CHIME at Redshifts z=1 and 2 via the IllustrisTNG Simulations}

\author[Polzin et al.]{
Ava Polzin$^{1,2}$\thanks{E-mail: apolzin@uchicago.edu},
Laura Newburgh$^{3}$,
Priyamvada Natarajan$^{2,3,4}$, 
and Hsiao-Wen Chen$^{1}$
\\
$^{1}$Department of Astronomy \& Astrophysics, The University of Chicago, Chicago, IL 60637, USA\\
$^{2}$Department of Astronomy, Yale University, New Haven, CT 06511, USA\\
$^{3}$Department of Physics, Yale University, New Haven, CT 06511, USA\\
$^{4}$Yale Center for Astronomy \& Astrophysics, Yale University, New Haven, CT 06520, USA
}

\date{Accepted XXX. Received YYY; in original form ZZZ}

\pubyear{2024}

\begin{document}
\label{firstpage}
\pagerange{\pageref{firstpage}--\pageref{lastpage}}
\maketitle

\begin{abstract}
 
The Canadian Hydrogen Intensity Mapping Experiment (CHIME) is a drift-scan interferometer designed to map the entire northern sky every 24 hours. The all-sky coverage and sensitivity to neutral hydrogen flux at intermediate redshifts makes the instrument a resource for other exciting science in addition to cosmology for which it was originally designed. Here, we demonstrate its utility for the study of the  \ion{H}{i}  content of galaxy populations across environments and redshifts. We use simulated data from the IllustrisTNG project to generate mock CHIME-like intensity maps, which we cross-correlate with various tracers -- including galaxies and galaxy clusters -- to recover aggregate \ion{H}{i} signals from stacking analyses. We find that there is more flux in stacks on galaxy clusters or cluster member galaxies compared to those on a general galaxy catalog due to the enhanced number of \ion{H}{i}-rich sources included in the CHIME primary beam. We report that it is possible to infer an average $M_\mathrm{HI}$ for clusters as a function of redshift and selection criteria from the signal in their averaged stacks despite the instrument's low spatial resolution. This proof-of-concept result opens up a promising, and timely, new avenue to measure the evolution of the neutral hydrogen content in intermediate-to-high redshift galaxy clusters via cross-correlation of galaxy cluster catalogs with \tcm intensity maps.
\end{abstract}


\section{Introduction}

The Canadian Hydrogen Intensity Mapping Experiment (CHIME; \citealt{2022ApJS..261...29C}) is a cylindrical drift-scan interferometer operating between 400-800\,MHz at the Dominion Radio Astrophysical Observatory in Penticton, Canada. CHIME was designed to use the redshifted \tcm emission of neutral hydrogen to measure large scale structure via a low spatial resolution, high spectral resolution survey (referred to as intensity mapping; \citealt{2004MNRAS.355.1339B, 2008PhRvL.100i1303C, 2008MNRAS.383..606W, 2009astro2010S.234P, 2015PhRvD..91h3514S}) at frequencies corresponding to $2.5 > z > 0.8$. To this end, maps from CHIME cover $\sim$60 per cent of the Northern sky daily with sub-degree angular resolution ($0.52^\circ-0.26^\circ$ for the synthesized beam), which translates to physical scales of $50-10$ Mpc \citep{2014SPIE.9145E..4VN, 2022ApJS..261...29C}.

Although the $2.5 > z > 0.8$ redshift range was selected for CHIME observations to constrain models of Dark Energy, it also corresponds to a rich era in galaxy evolution when the overall star formation efficiency is at its peak (`Cosmic Noon'; e.g., \citealt{2014ARAA..52..415M}). At these redshifts, the number density of quenched galaxies, or galaxies in which star formation has ceased, increases dramatically \citep[e.g.][]{2007ApJ...665..265F}. Since the abundance of neutral hydrogen in galaxies is a robust indicator of the existing reservoir of star-forming gas content, instruments like CHIME are extremely valuable 
for looking at trends in this reservoir at these specific cosmic times, and could possibly be used to quantify the average  \ion{H}{i}  mass of a target population from stacks.

CHIME's redshift band and sensitivity to neutral hydrogen could provide a new probe of the integrated  \ion{H}{i}  content of galaxy clusters spanning the earliest assembling clusters around $z\sim2$ to the quenching of star formation around $z\sim$1. At high redshift, galaxy clusters are generally thought to be gas-rich and contain large reservoirs of neutral hydrogen \citep[e.g.,][]{2001PASA...18...64F, 2004MNRAS.355.1339B}. 
There are ongoing and planned experiments aimed at studying the evolution of cluster \ion{H}{i} in individual galaxy clusters. The Five-hundred-meter Aperture Spherical radio Telescope \citep[FAST,][]{2019SCPMA..6259502J, 2020RAA....20...64J, 2020Innov...100053Q} should be able to measure the \ion{H}{i} signal from individual neutral hydrogen-rich galaxy clusters out to $z \sim 1.5$~\citep{2017RAA....17..101A}. However, even the best-case projected integration times for FAST to reach the required depth are on the order of hours to days, depending upon cluster redshift, which limits potential survey scope. 
Data from CHIME can be used to better understand the average reservoir of neutral gas in dense environments at the population level. Such measurements potentially have the power to statistically distinguish the role of AGN feedback in driving the evolution of clusters' neutral hydrogen reservoir and confirm whether it follows model predictions as a function of redshift \citep[e.g.,][]{2016MNRAS.456.3553V, 2019MNRAS.483.3336T, 2023ApJ...945L..28D}.

Due to poor spatial resolution, experiments like CHIME cannot directly measure the flux from a galaxy, galaxy group, or galaxy cluster. Instead, data from CHIME has been stacked in voxels (RA, Dec, $z$) on known source locations from eBOSS \citep[][]{2016AJ....151...44D} to detect cosmological neutral hydrogen in emission line galaxies (ELGs), luminous red galaxies (LRGs), and quasars (QSOs) \citep[][]{2023ApJ...947...16A}, and has been correlated with Lyman-alpha emitters to detect a decrement of neutral hydrogen in absorber systems \citep{2024ApJ...963...23A}. These detections highlight that the CHIME data set in principle has access to the aggregate quantity of neutral hydrogen in sources at these intermediate redshifts.
Catalogs of hundreds of clusters are already available from optical, X-ray, and CMB surveys spanning this redshift range \citep[e.g.,][]{2015ApJS..216...27B, 2021MNRAS.500.1003W, 2021ApJS..253....3H, 2022MNRAS.513.3946W, bulbul2024srgerosita}, and with upcoming experiments like Simons Observatory \citep{2019JCAP...02..056A} and CMB-S4 \citep{2016arXiv161002743A} coupled with follow-up enabled by \textit{JWST}, Roman, and Lynx, we expect this number to grow substantially \citep{2019BAAS...51c.279M}, with thousands of clusters expected to be detected at $z\sim1$ and hundreds expected to be detected at $z \sim 2$ with CMB-S4.

In preparation for cross-correlating CHIME data with galaxy cluster (or other) catalogs to answer questions related to galaxy evolution, or, more specifically, the evolution of the gaseous medium in and around galaxies, it is important to understand which objects actually contribute to the measured  \ion{H}{i}  flux in stacked data. Our low spatial resolution ensures that we are able to measure the integrated flux from objects that fall in the beam; however, that low resolution comes with the caveat that we are not able to resolve individual objects and/or their immediate environments. Instead, structures surrounding our target objects are also likely to be included in the beam and in our stacking analyses. It is essential that we understand how much surrounding ancillary objects contaminate our stacked flux measurements. In this paper, we use the Illustris TNG300 suite of simulations of large scale structure and its neutral hydrogen content to estimate whether CHIME could be used to measure the  \ion{H}{i}  content in galaxy clusters across this interesting epoch. 

We assume the general cosmological parameters from \citet{2016AA...594A..13P} used by the TNG300 simulation \citep[][]{2018MNRAS.475..648P, 2018MNRAS.475..676S, 2018MNRAS.475..624N, 2018MNRAS.477.1206N, 2018MNRAS.480.5113M} of the IllustrisTNG project \citep[][]{2019ComAC...6....2N}, $h = 0.6774$, $\Omega_M = 0.3089$, and $\Omega_\Lambda = 0.6911$, and we adopt the  \ion{H}{i}  -- \textsc{H$_2$} transition prescription from \citet[][]{2011ApJ...728...88G} to assign  \ion{H}{i}  masses to each halo.

The structure of this paper is as follows: in Section~\ref{sec:sim}, we discuss the simulated data that facilitates this analysis; in Section~\ref{sec:stacking}, we go through the steps of our stacking analysis; in Section~\ref{sec:mass}, we lay out an example means of recovering $M_\mathrm{HI}$ from CHIME galaxy cluster stacks; and in Section~\ref{sec:discuss}, we consider the broad applicability of this work.

\section{Simulating CHIME observations} \label{sec:sim}
In this section we describe how we generate a mock CHIME observation of an IllustrisTNG simulation field. This paper is meant to describe a proof-of-concept reconstruction, and so we do not follow the simulation template procedure outlined in \citet{2023ApJ...947...16A}, which was used to find parameter errors on model parameters such as the 21~cm power spectrum ($P_{\mathrm{HI},g}$), the bias of galaxy and 21~cm fields ($b_g$, $b_\mathrm{HI}$), etc., and included effects such as fingers-of-god, nor do we include processing steps such as filtering and weighting. Instead, we adopt a less rigorous approach of convolving projections of the TNG300 volume with the synthesized CHIME beam shape and using that as a proxy for the stacking measurement to understand whether CHIME can in principle recover  \ion{H}{i}  information from high redshift clusters.

\subsection{TNG300 simulation from the IllustrisTNG project}

We select the TNG300 simulation, part of the Illustris suite of cosmological simulations,
as the parent simulation for generating our mock CHIME maps due its large volume, native inclusion of  \ion{H}{i}, and relevant redshift coverage.
In the TNG300 simulation, the evolution of the gas content - both molecular and atomic hydrogen - is tracked for every sufficiently massive ($M_\mathrm{star} \ge 5 \times 10^{10} M_\odot$ or $M_\mathrm{gas} \ge 10^{9} M_\odot$) subhalo across
environments. 

The TNG300 \textit{Molecular and atomic hydrogen (\textsc{HI+H2}) galaxy contents} catalogs \citep[][]{2018ApJS..238...33D, 2019MNRAS.487.1529D} provide the neutral hydrogen mass of each subhalo at two discrete redshifts, $z=1$ and $z=2$, that also overlap with the CHIME observation band. Since $z=1.5$ (568 MHz) falls into the radio frequency interference-afflicted $500-585$ MHz sub-band and is inaccessible to us in analysis of real CHIME data, we do not include it in our mock data analysis either. We select the \citet[][alternately referred to as \citetalias{2011ApJ...728...88G}]{2011ApJ...728...88G} \ion{H}{i} model from those included in the \textit{Molecular and atomic hydrogen (\textsc{HI+H2}) galaxy contents} catalogs; the similarity of TNG300  \ion{H}{i}  abundances to observations is discussed further in \citet{2019MNRAS.487.1529D}.

\label{app:gk}

With the tabulated neutral hydrogen masses for each sufficiently massive subhalo from catalogs at $z=1$ (710 MHz) and at $z=2$ (473 MHz), we construct mock high-resolution maps, which are then smoothed with the CHIME synthesized beam to simulate observational data.

\subsection{Constructing a beam-like convolutional kernel}

To generate the low-resolution CHIME intensity maps from high-resolution TNG300 simulations, we require a frequency-dependent synthesized beam model of CHIME. The cylindrical design of the CHIME reflector has a different cross-sectional beam profile in the North-South (N-S) and East-West (E-W) directions.

We assume an approximate form of the N-S beam shape, inspired by the triangular window function that resulted from the baseline-dependent inverse variance weighting scheme described in \citet{2023ApJ...947...16A}. The triangular window function is given by:
\begin{equation}
    W(y) = \frac{1 - |y/y_\mathrm{max}|}{\sum{1 - |y/y_\mathrm{max}|}}
\end{equation}
where $y$ is the physical distance along the North-South direction in units of wavelength and $y_\mathrm{max}$ is the longest North-South baseline in units of wavelength. The N-S beam model is the Fourier transform of this window function. As is standard in radio interferometry, this produces baseline-dependent, frequency-dependent scale on the sky.

We use the synthesized E-W beam model from \citet{2023ApJ...947...16A} at 710\,MHz ($z=1$), with a few simplifications: we have limited the beam to hour angles less than $\pm 4^{\circ}$ which neglects the effects of aliasing\textbf{; t}his beam is computed by beamforming to declination $\delta=0^{\circ}$, and we use only the \textit{Y}-polarized beam and include only inter-cylinder baselines (baselines that span more than one cylinder). The full form of the synthesized beam is described in more detail in Section~4.3.4 of \citet{2023ApJ...947...16A}. For the E-W beam at 473\,MHz ($z=2$), we use full width at half maximum (FWHM) $\propto 1/\nu$ to re-scale the 710 MHz ($z=1$) beam model. We also interpolate the beam function to a coarser grid-size to convolve more easily with the simulations.

\begin{figure}
	\includegraphics[width=\columnwidth]{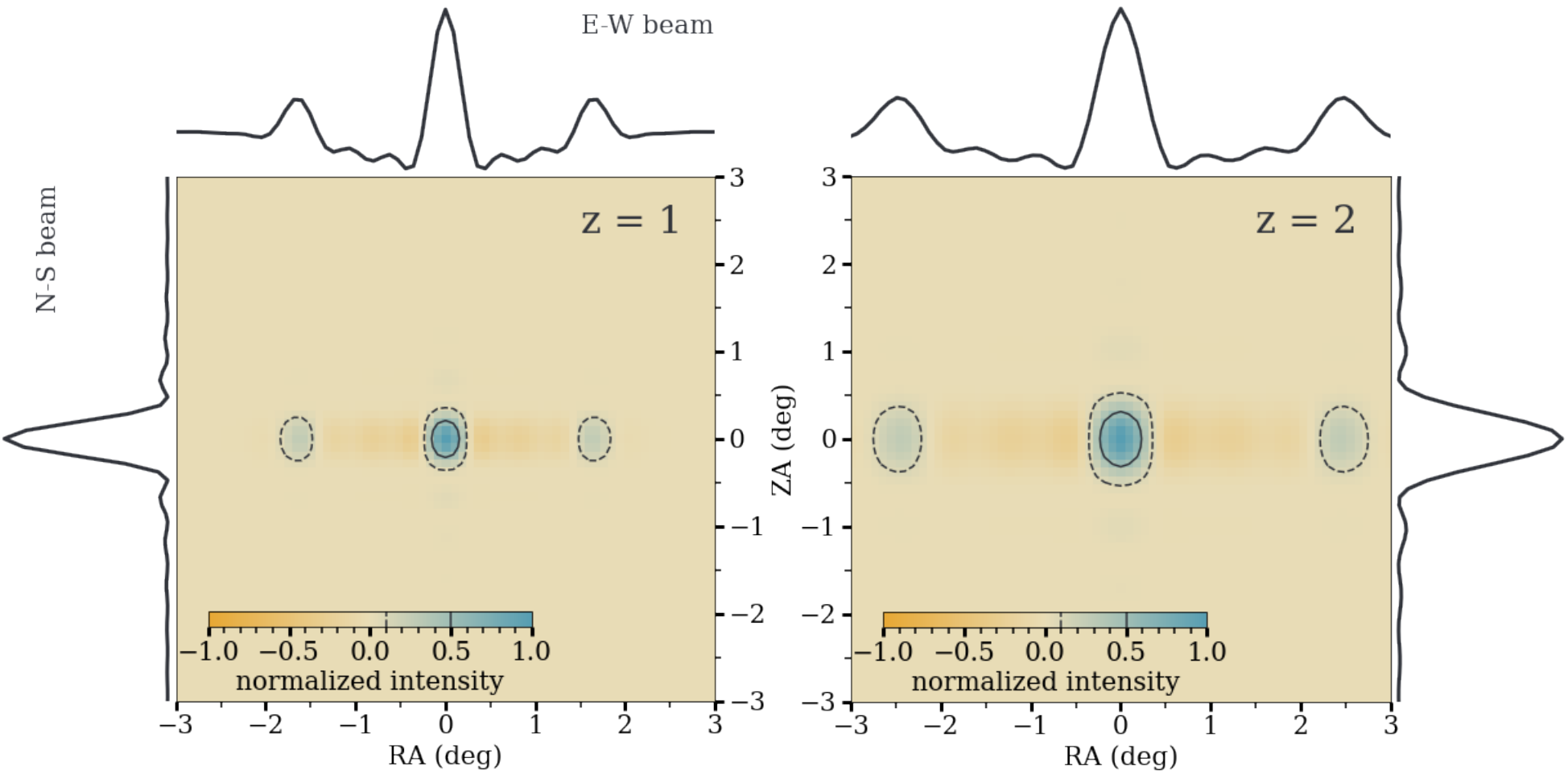}
    \caption{
    We show the cross section of synthetized beam models used in this paper along the the E-W (x axis) and N-S (y axis) direction, sampled at the resolution of the stacked CHIME maps ($5\farcm3$ in RA and $5\farcm7$ in zenith angle). The synthesized beam models are used as convolutional kernels to go from a TNG300 gas mass map to a CHIME-like intensity map. \emph{Left}: Shown for the the 710\,MHz ($z = 1$) synthesized beam model, which is the outer product of the E-W and N-S models. \emph{Right}: Shown for the 473\,MHz ($z = 2$) synthesized beam model.
    }
    \label{fig:beam}
\end{figure}

The two-dimensional, synthesized beam is then the outer product of the N-S and E-W cross-sections. Although this does not capture the rich structure of the CHIME beam \citep{2022ApJ...932..100A}, the simplified beam is sufficient for the explorations in this paper as we focus here on the polarization-independent flux measured in the primary component of the synthesized beam. These beam models, both synthesized and cross-sectional, are shown in Fig. \ref{fig:beam}.

\subsection{Generating mock intensity maps}

Here we outline the procedure adopted to generate mock intensity maps from the simulated data. Assuming that the  \ion{H}{i}  gas is optically thin in the sources we are examining, we use the formalism from \citet{1952ApJ...115..206W} to derive the neutral hydrogen intensity from  \ion{H}{i}  mass \citep{2011piim.book.....D}:
\begin{equation}
    \label{eq:draine}
    \left(\frac{F_\mathrm{obs}}{\textnormal{Jy MHz}}\right) = 2.022\times10^{-8} \, \left(\frac{M_{\textsc{HI}}}{\textnormal{M}_\odot}\right) \, \left(\frac{d_L}{\textnormal{Mpc}}\right)^{-2}
\end{equation}
The observed flux, $F_\mathrm{obs}$, is defined as the flux density integrated over the frequency range of the relevant CHIME sub-band. This flux should be equivalent to Stokes \textit{I}, which combines the two linear polarizations for total intensity ($I=X^2 +Y^2$); for the purposes of this paper, we use the CHIME measured \textit{Y} polarization beam as a proxy for the Stokes \textit{I} beam due to its well-understood behavior in these two redshift regimes, noting that the two instrumental polarizations are shown to return consistent results in \citet[][]{2023ApJ...947...16A}, so that any difference in the \textit{X}- and \textit{Y}-polarized beams should have a negligible impact on the analysis in this work.

The luminosity distance, $d_L$, is defined consistent with our chosen cosmology and the simulation volume -- for $z = 1$, $d_L$ = $6.8\times10^3$ to $7.0\times10^3$ Mpc and for $z = 2$, $d_L$ = $1.59 \times 10^4$ to $1.60 \times 10^4$ Mpc. 
We convert to flux density in Jy by dividing off the width of the CHIME frequency sub-band (0.390 MHz), which assumes that we know the spectral position of targets to within $dz \lesssim 0.001$ at $z = 1$ and $dz \lesssim 0.002$ at $z = 2$.  In stacking, see Section \ref{sec:stacking}, we assume that all objects in the mock intensity map are at $z = 1 \pm dz$ or 2 $\pm dz$. Selecting a different range in redshift/frequency will impact the noise properties of the stacks as well as the observed flux. Foreground filtering strongly attenuates the signal with increasing bandwidth. To approximate this effect in our mock observations, we apply a corrective factor of 0.2 to the flux, corresponding to the mean foreground filtering attenuation between catalogs used in \citet{2023ApJ...947...16A}. This is for our narrowest CHIME band, comparable to the precision of a spectroscopic catalog. More advanced foreground filtering may allow for a greater tolerance in $dz$ in the future.

We deposit the flux from each TNG300 subhalo onto a grid matching the angular scale of CHIME map pixels ($5\farcm3$ in RA and $5\farcm7$ in Dec). The CHIME hybrid beamformed maps have a zenith angle-dependent pixel size; here we choose the highest resolution pixel scale ($5\farcm3 \times 5\farcm7$) corresponding to ZA = 0$\degr$. Included  \ion{H}{i}  sources will be further smoothed by later convolution with the low angular resolution CHIME beam.

Given the limited angular extent of each TNG300 slice at $z = 1$ and $z = 2$, we tile rotated (and flipped) versions of the same projected map to create a larger map to smooth and stack. Since we also permute the projected maps from the existing (300 Mpc)$^3$ simulation volume with respect to the reference position used in calculating $F_\mathrm{obs}$, the final tiled map encompasses 384 times the area of a single projected simulation slice, totalling $\sim 4100$ deg$^2$ at $z = 1$ and $\sim 1900$ deg$^2$ at $z = 2$. 
(See Appendix \ref{app:makemap} for more details.)  It is this larger intensity map that is then convolved with the CHIME beam model. The mock sky coverage is comparable to the extent of the eBOSS fields used in \citet{2023ApJ...947...16A}. We add Gaussian noise with $\sigma = 0.6$ mJy \citep{2023ApJ...947...16A} to the beam-convolved map in order to better replicate CHIME observations. 

\begin{figure*}
	\includegraphics[width=2\columnwidth]{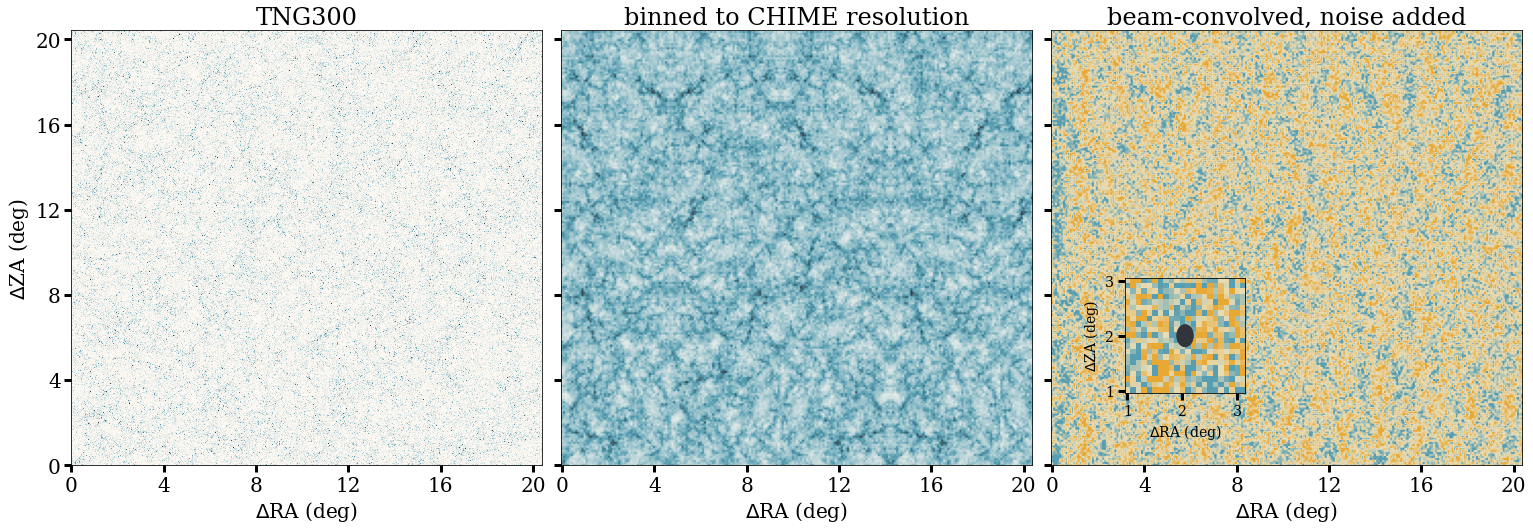}
    \caption{An example zoom-in region of a neutral hydrogen map at different points in processing. The leftmost panel offers a higher resolution look at the spatial distribution of TNG300 galaxies as mapped by their normalized  \ion{H}{i}  flux. The middle panel shows those same galaxies binned to reflect the spatial resolution of CHIME. The rightmost panel shows the middle panel convolved with the CHIME beam (see Fig. \ref{fig:beam}) with CHIME-like noise added. The inset panel shows a zoomed-in region of the the beam-convolved map overlaid with an ellipse representing CHIME's primary beam FWHM. The portion of the map used here is 4 per cent of the full $z = 1$ tile that is used in our mock stacking analyses. 
    }
    \label{fig:procex}
\end{figure*}

Fig. \ref{fig:procex} shows the progression from the higher resolution native TNG300 mock intensity map to the beam-smoothed mock intensity map with realistic noise. We note that we do not include foreground emission, frequency masking from radio frequency interference, or point source removal. Instead, the results we present provide a proof-of-principle extraction of astrophysical neutral hydrogen in clusters from intensity mapping experiments like CHIME.

\section{Mock stacking analyses} \label{sec:stacking}

\begin{figure*}
	\includegraphics[width=2\columnwidth]{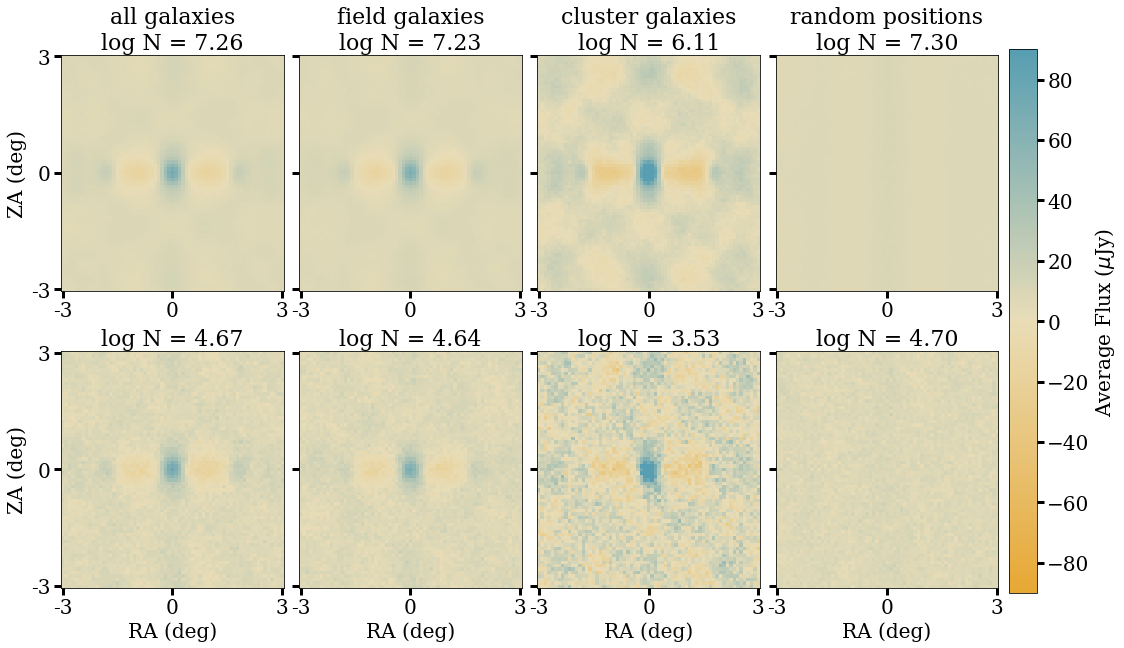}
    \caption{Output stacks at $z = 1$ for different galaxy populations ($M_\mathrm{star} \ge 10^{10} \, M_\odot$). To demonstrate that the signal we are measuring is real, we also stack on randomly generated positions (\emph{rightmost panels}). \emph{Top:} The ``true'' stacks from TNG300, using all of the massive galaxies in simulation volume. \emph{Bottom:} Stacks based on $N$ randomly sample massive galaxies from the full catalogs. This less well-sampled \textit{all galaxies} stack is directly comparable to the weighted eBOSS/QSO stack from \citet{2023ApJ...947...16A}. The galaxy stacks are then sliced to make the $z = 1$ panels of Fig. \ref{fig:galstacks}.}
    \label{fig:stackex}
\end{figure*}

We stack the signal from the simulated  \ion{H}{i}  maps for the two redshifts separately. For each source in the simulation, we create a two-dimensional, $6^\circ \times 6^\circ$ cutout in RA and Dec, following Section~4.7 of \citet{2023ApJ...947...16A}. These cutouts are then directly co-added, returning a `stacked' signal from the average of the co-added signals (see Fig. \ref{fig:stackex} for an example). The intensity map area of our $z = 1$ mock is comparable to the area used in the eBOSS/LRG and QSO cross-matched stacks from \citet{2023ApJ...947...16A}, while the area of our $z = 2$ mock is comparable to the area used in the eBOSS/ELG cross-matched stack. We do not expect that the $z$-dependent difference in area impacts our ability to compare mock stacking analyses across redshifts, given that the signal we are measuring is innately normalized by virtue of averaging the stacks. 

In order to isolate the contribution to the  \ion{H}{i}  signal from galaxy clusters and their constituent galaxies, we create separate stacks based on our cluster population-specific maps at each redshift in addition to the full map used for much of the analysis: mass-selected galaxy clusters only and radius-selected galaxy clusters only. The specific maps and stacks are as follows:

\begin{enumerate}
    \item \textit{all galaxies}; stacked on all galaxies\footnote{This first stack is intended to be more easily compared with the analysis done in \citet[][]{2023ApJ...947...16A}.}
    \item \textit{field galaxies}; stacked on field galaxies
    \item \textit{cluster galaxies}; stacked on cluster galaxies
    \item \textit{all clusters}; stacked on all galaxy clusters with the centre of the stack defined by the centre of mass of the halo.
    \item \textit{mass-selected clusters}; stacked on mass-selected galaxy clusters with haloes with a mass $\ge 10^{14} \, M_\odot$ at $z=0$ as traced by central galaxy membership at $z = 0, 1, $ and 2. 
    \item \textit{radius-selected clusters}; stacked on radius-selected galaxy clusters with haloes with comoving $R_{200} > 0.75 \, h^{-1}$ Mpc, independently determined for $z = 1$ and $z = 2$.
    \item \textit{random positions}; stacked on $N$ random locations. 
\end{enumerate}

The two cluster selection criteria are intended to ensure that we are accounting for proto-clusters that should be present at $z = 2$ as expected in the $\Lambda$CDM cosmology (mock catalog construction is detailed further in Appendix \ref{appA}).

We define field galaxies as subhaloes which do not belong to a group designated as a cluster (due to either mass or radius selection).

Though all galaxies down to TNG300's mass resolution limit for the \textit{Molecular and atomic hydrogen galaxy contents} catalog ($M_\mathrm{star} \ge 5 \times 10^{10} M_\odot$ or $M_\mathrm{gas} \ge 10^{9} M_\odot$) are included in the maps, those maps are only stacked on galaxies with $M_\mathrm{star} \ge 10^{10} M_\odot$ (as defined by the \textit{SubhaloStellarPhotometricsMassInRad} group catalog field); this roughly corresponds to the lowest mass bin in the eBOSS ELG sample \citep[e.g.][]{2019ApJ...871..147G}.

\begin{figure*}
	\includegraphics[width = 2
\columnwidth]{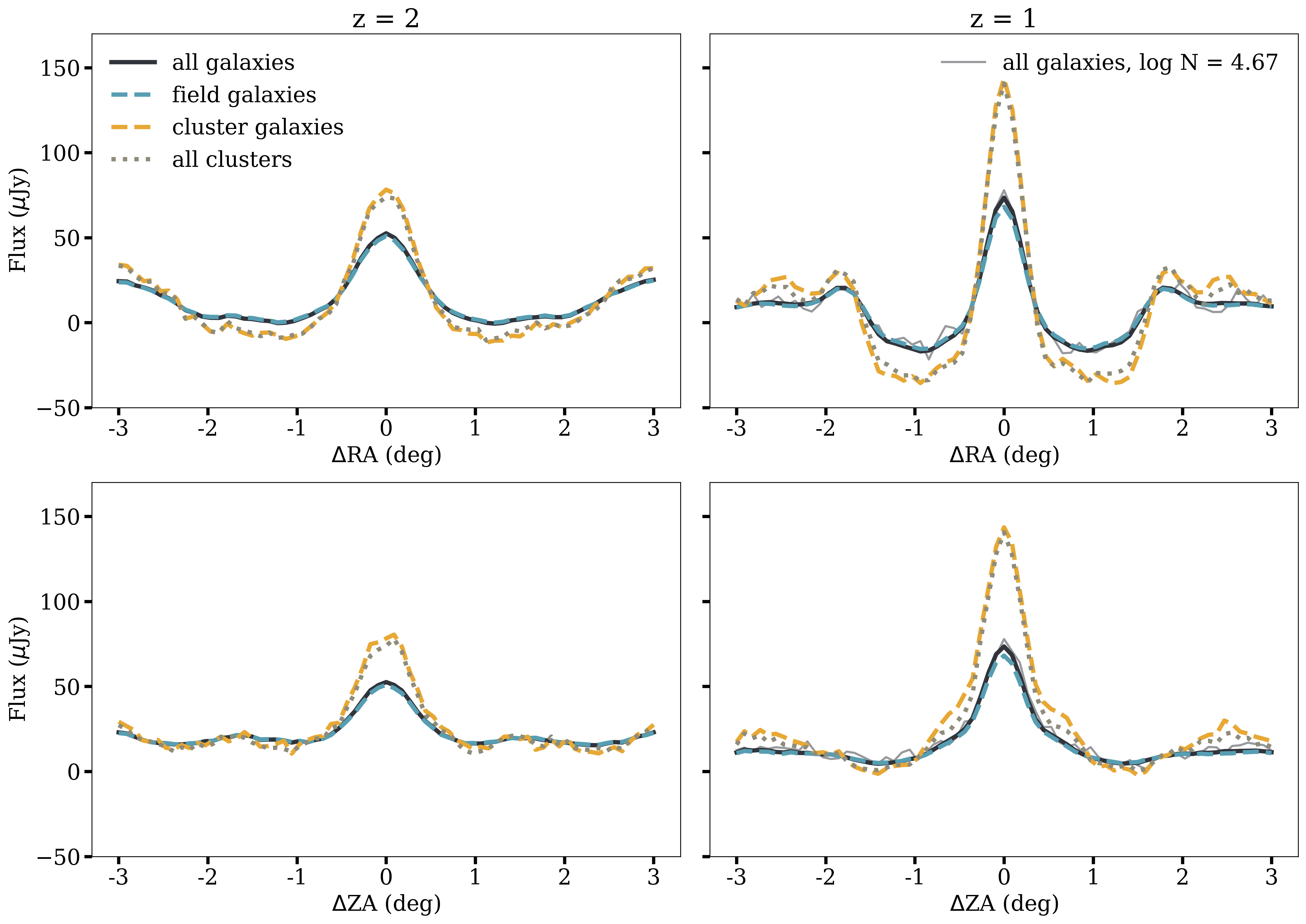}
    \caption{The results of our stacking, as described in Section \ref{sec:stacking}, on the locations of all individual galaxies/subhaloes in the TNG300 volume that meet our minimum $M_\mathrm{star} \ge 10^{10}\, M_\odot$ threshold. The upper panel shows the stacks sliced along the central pixel in ZA, and the bottom panel shows the stacks sliced along the central pixel in RA. At left, we show the $z = 2$ stacks and at right, we show the $z = 1$ stacks. The disparity between stacks on field and cluster galaxies is likely due to more galaxies in the crowded cluster environments falling into the CHIME beam. That is, stacking on cluster galaxies returns the  \ion{H}{i}  signal from the galaxy clusters themselves. This is shown by the dotted ``all clusters'' curve matching the flux returned by stacking on cluster galaxies. Alongside the full $z = 1$ stacks, we also show a less well-sampled \textit{all galaxies} stack that is directly comparable to the weighted eBOSS/QSO stack from \citet{2023ApJ...947...16A}. It is apparent that the measured peak flux is the same between this stack and the ``true'' \textit{all galaxies} stack, while the noise properties are somewhat different.}
    \label{fig:galstacks}
\end{figure*}

\begin{figure*}
	\includegraphics[width = 2
\columnwidth]{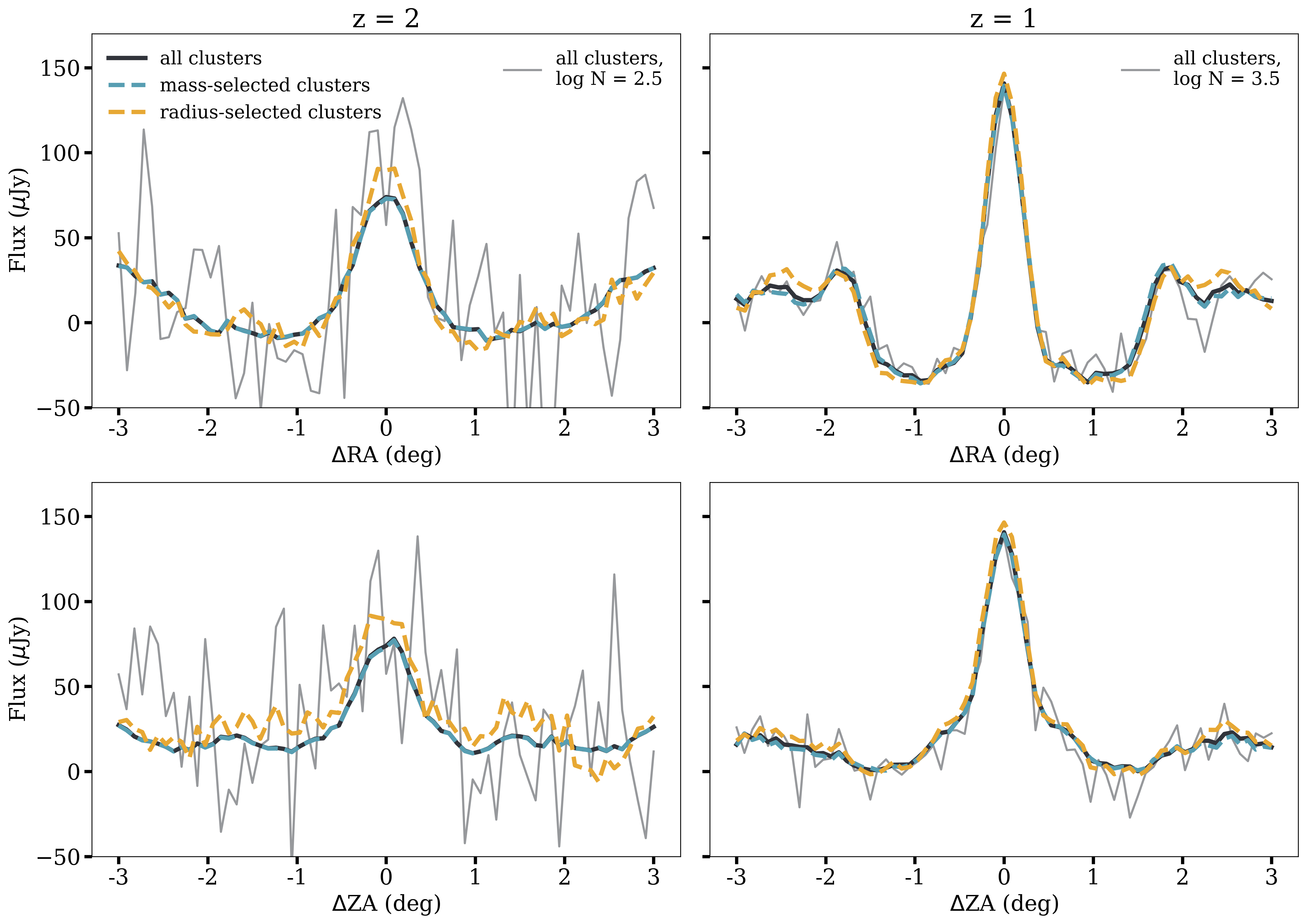}
    \caption{The results of our stacking, as described in Section \ref{sec:stacking}, on the locations of cluster-identified haloes. The upper panel shows the stacks sliced along the central pixel in ZA, and the bottom panel shows the stacks sliced along the central pixel in RA. At left, we show the $z = 2$ stacks and at right, we show the $z = 1$ stacks. Solid and dashed lines show results for the full map stacked on galaxy cluster locations. At $z = 2$ ($z = 1$), the black curve shows the stack on 106752 (118656) clusters, the blue curve shows the stack on 105984 (106368) mass-selected clusters, and the yellow curve shows the stack on 9984 (61440) radius-selected clusters in the full CHIME-like map.
    In addition to the ``true'' stacks based on all relevant sources in the TNG300 volume, we underplot example cross-sections from stacks based on 300 $z= 2$ and 3000 $z = 1$ clusters randomly selected from the full catalog (shown in gray). These stacks are consistent with the number of clusters we anticipate will be in catalogs facilitated by next-generation instruments. Notably, the shape and amplitude of the cross-sections are preserved despite the increased noise due to the small sample sizes.}
    \label{fig:clusterstacks}
\end{figure*}

We show example stacks in Fig. \ref{fig:stackex}. The top row includes stacks (i) - (iii) and (vii) on the full catalog returned for the simulation volume, and the bottom row shows the same stacks for a more realistic number of randomly selected galaxies in the simulation volume. These realistic stacks are directly comparable to those in \citet{2023ApJ...947...16A}, including the same number of target galaxies. The amplitude of the stacked signal is equivalent between the full catalog and realistic stacks, with the noise properties varying as predicted based on the number of targets used in each stack. The implications of stacking on unresolved target source is described in Section \ref{sec:mass}.

\section{Results} \label{sec:mass}

\subsection{Neutral hydrogen environment}

The results of stacks (i) - (iii) are shown in Fig. \ref{fig:galstacks}, where it is apparent that the measured average neutral hydrogen flux is higher from galaxies associated with clusters than from field galaxies. 
We found this was explained by the crowded cluster environment causing 'source confusion': the CHIME beam is large enough to encapsulate all galaxies in the cluster\footnote{The primary beam has a resolution of $\approx10-20$ Mpc, while clusters are $\approx0.5-1$ Mpc in radius in this redshift range.}, such that we are adding the flux from all galaxies in the cluster each time we stack on a single galaxy in the cluster. This means we measure the aggregate  \ion{H}{i}  emission from all galaxies, not the average  \ion{H}{i}  per galaxy, in these stacks. As a result, environments with more galaxies have a higher stacked signal than environments with fewer galaxies (such as the field galaxies). This is consistent with Fig. \ref{fig:clusterstacks}, which shows the results for stacks (iv) - (vi).

Fig. \ref{fig:galstacks} shows the results from stacking on all galaxies in the sample at a redshift $z=1$. That simulation is comparable in number of galaxies and mean redshift to the published CHIME results for the QSO and ELG samples. We find that the flux associated with the \textit{all galaxies} sample indeed has a flux peak value within a factor of a few of the CHIME results from stacking on the eBOSS/QSO catalog with median $z \sim 1.2$, which is enough consistency to not impact the conclusions drawn in this paper. The disparity between the \citet{2023ApJ...947...16A} peak flux and the one measured here may come from differences in approach between simulations and real data, such as the potential for increased background flux in our convolved maps due to flattening over the entire simulation volume. In addition, we recover the shape of the co-added signal, whose negative value on each side of the peak was shown in the CHIME data and attributed to the synthesized beam which we also use in this paper.

We confirm that the cluster galaxy flux traces galaxy cluster  \ion{H}{i}  by overplotting the cluster stacks on the cluster-member galaxy stacks in Fig. \ref{fig:galstacks} where it can be seen that the stack on cluster galaxies matches the stack from cluster centre positions only.
As noted above, this is a consequence of measuring the aggregate emission from all sources in the beam in the stacks.
Conversely, the stack on field galaxies have lower measured average flux because the lower number counts mean there is less aggregate  \ion{H}{i}  emission from that region.
The field galaxy flux is comparable to the \textit{all galaxies} flux. Because the sample size of field galaxies is $\sim100\times$ larger than the sample size of cluster galaxies, the \textit{all galaxies} signal is dominated by the field galaxies.

This same statistical effect is driving the similarity of the mass-selected cluster stack and the \textit{all cluster} stack in Fig. \ref{fig:clusterstacks} -- at $z = 2$, 99 per cent of clusters are included in the mass-selected sample and at $z = 1$, 90 per cent of clusters are in the mass-selected sample.
The very different flux between radius-selected clusters (which are selected for their properties at $z = 1$ and 2 respectively) and mass-selected clusters (selected for their properties at $z = 0$) at $z = 2$ is indicative of the fact that radius-selected clusters are intrinsically larger and more assembled on average than mass-selected clusters at the same redshift and so radius-selected clusters are denser and will have more aggregate  \ion{H}{i}  emission in the CHIME beam than the mass-selected cluster.
The mass-selection criterion is intended to include proto-clusters in our analysis, as a result they are smaller and contain less \ion{H}{i}. Thus, it is possible to look at the relative  \ion{H}{i}  content of different populations of clusters (categorized based on their initial detection method or other properties) even without fully understanding an instrument's flux calibration. This opens up opportunities to look at the ratio of average  \ion{H}{i}  content between populations. If the frequency-dependent flux calibraton is determined, this can also allow comparisons between redshifts.

\subsection{Mass recovery for complete TNG300 catalogs}

\begin{figure}
	\includegraphics[width=\columnwidth]{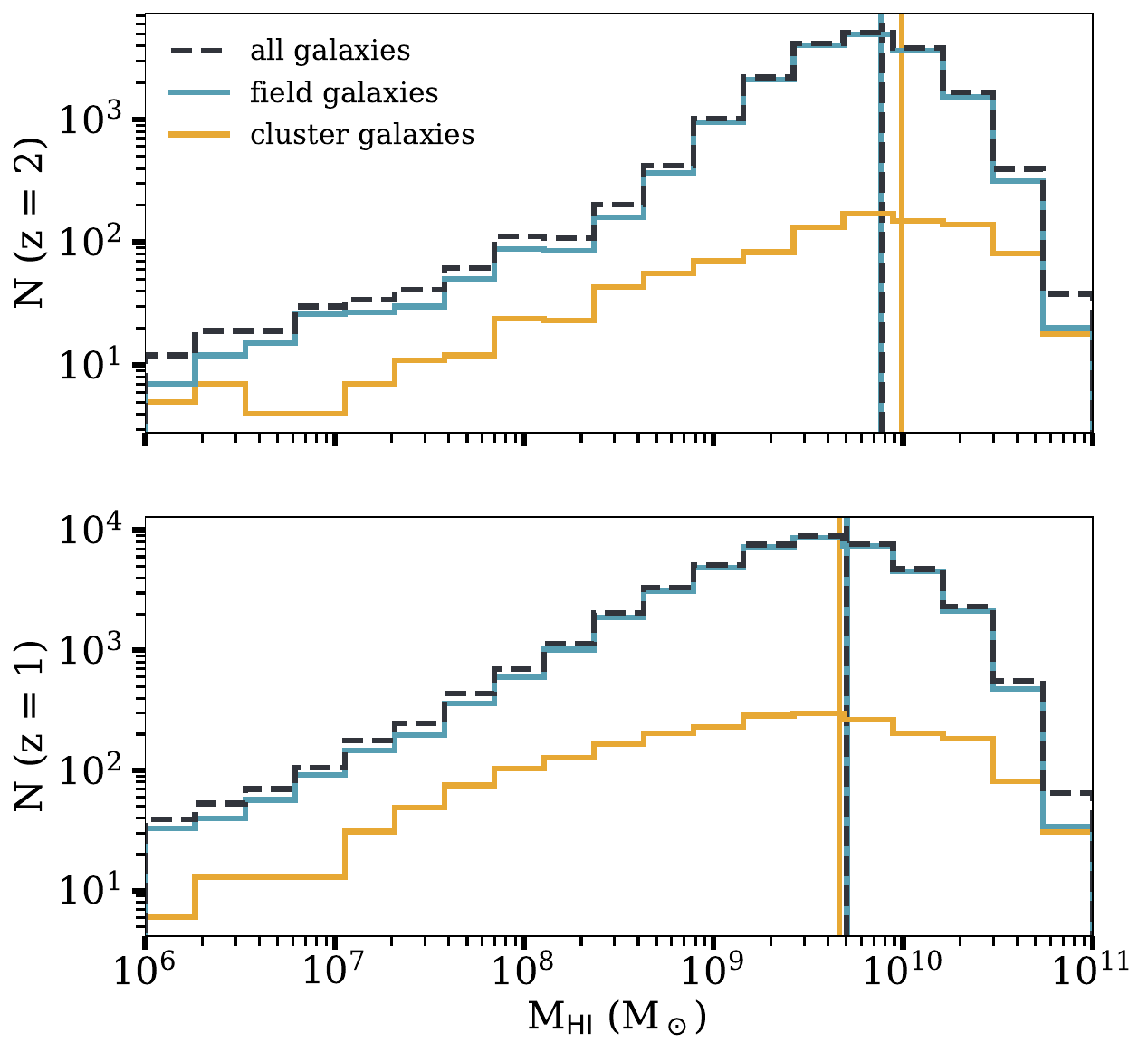}
    \caption{The intrinsic neutral hydrogen mass distribution from TNG300 in each of three subhalo populations  -- all galaxies (subhaloes), field galaxies, and cluster member galaxies -- for the galaxies included in the stacks at $z = 2$ (top) and $z = 1$~(bottom). The mean \ion{H}{i} mass for each population is shown as a vertical line. As with other analyses in this paper, we limit each sample to only include galaxies with $M_\mathrm{star} \ge 10^{10}\, M_\odot$, in-keeping with the stellar mass limit of the eBOSS galaxies. Note that the overall normalization changes between the top and bottom panels.}
    \label{fig:galmass}
\end{figure}

\begin{figure}
	\includegraphics[width=\columnwidth]{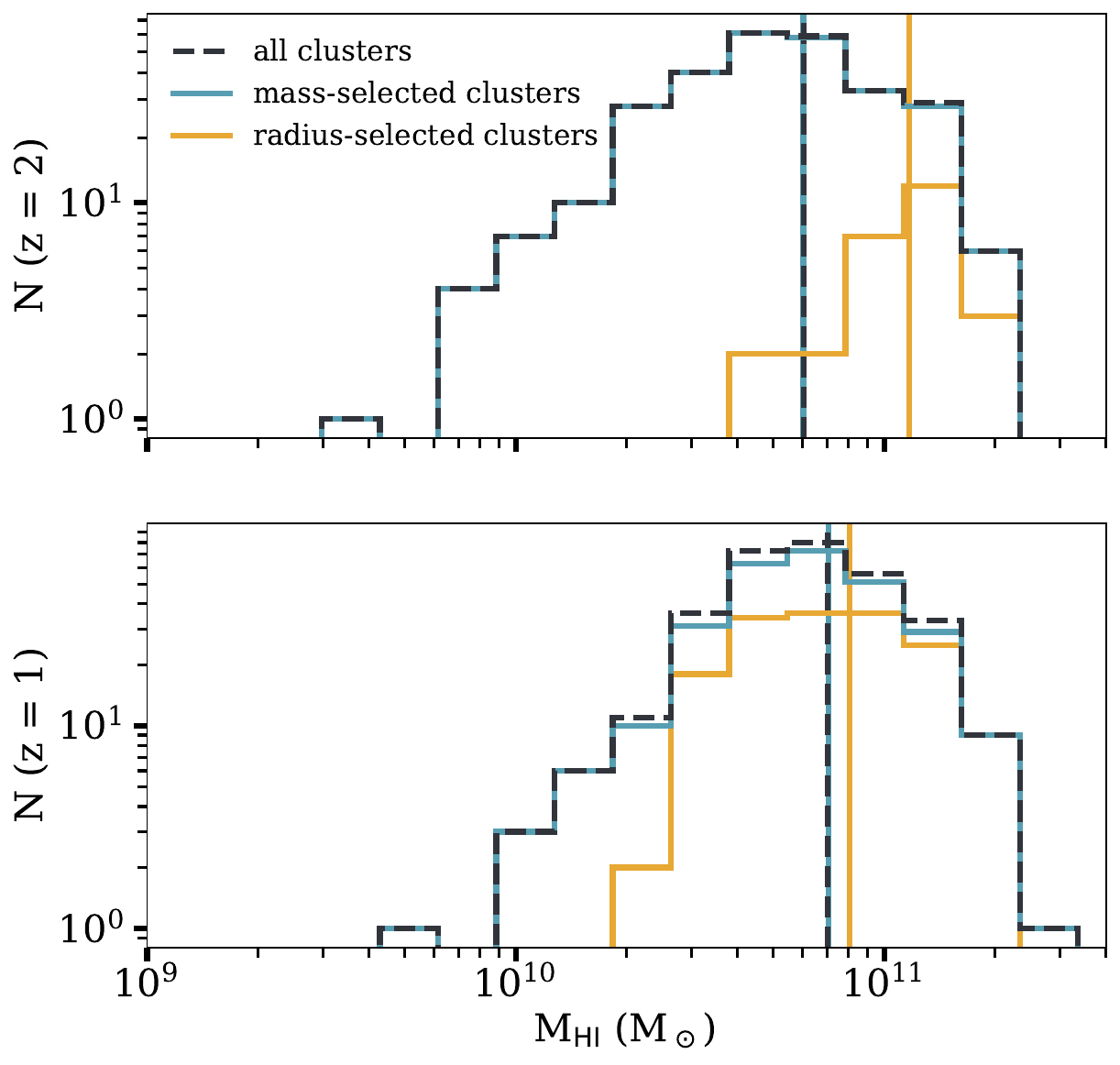}
    \caption{The intrinsic total neutral hydrogen mass distribution from TNG300 in each of three halo populations -- all clusters (haloes), mass-selected galaxy clusters, and radius-selected galaxy clusters -- for the clusters included in the stacks at $z = 2$ (top) and $z = 1$~(bottom). The mean \ion{H}{i} mass for each population is shown as a vertical line.}
    \label{fig:clustermass}
\end{figure}

The quantity of neutral hydrogen in the intergalactic medium directly informs the evolution of the galaxies in this environment. We can use this mock CHIME analysis to study whether we can use  \ion{H}{i}  flux measurements from galaxy clusters to infer their atomic neutral hydrogen mass at intermediate redshifts. As a test of this application, we compare the  \ion{H}{i}  mass inferred from the stacked signal within CHIME's primary beam and the mean  \ion{H}{i}  mass of the subhaloes (Fig. \ref{fig:galmass}) and haloes (Fig. \ref{fig:clustermass}) that were used in the construction of our mock intensity maps. 

For galaxy clusters, the flux in the averaged stacks exhibit similar trends to the mean  \ion{H}{i}  mass of the included haloes (i.e., that mass-selected and radius-selected clusters have approximately the same signal at $z = 1$, while at $z= 2$, radius-selected clusters have higher stacked flux and higher mean \ion{H}{i}; Fig. \ref{fig:clusterstacks} and Fig. \ref{fig:clustermass}), thus allowing us to estimate the neutral hydrogen mass from the stacked data. 
Using even a rough conversion to infer the mass, we find that this can be accomplished by inverting equation (\ref{eq:draine}) and accounting for the differences in spatial coverage at $z = 1$ and 2:
\begin{equation}
    \frac{M_\mathrm{HI}}{M_\odot} \approx \frac{4.945\times10^7}{C(z)}\, \left(\frac{F_\mathrm{pk,sub}}{\mathrm{Jy \, MHz}}\right) \left(\frac{\langle d_L\rangle}{\mathrm{Mpc}}\right)^2
\end{equation}
where $\langle d_L \rangle$ is the average luminosity distance of the stacking targets, $F_\mathrm{pk, sub}$ is the flux measured in the central pixel of a background-subtracted stack, and $C$ is a single redshift-dependent conversion factor. We generate a background stack to subtract by stacking on $N$ random locations. For the purposes of our calibration here, we select $N$ equal to the length of the catalog from which the original stack was created, and note that choosing too small a number may result in the inferred mass changing with repeated measurements. To consistently measure $F_\mathrm{pk, sub}$ for stacks with very different noise properties (primarily defined by the number of targets), we fit a Gaussian profile
\begin{equation}
    F(x) = n \,\exp(-0.5 \left(\frac{x - \mu}{\sigma_z}\right)^2)
\end{equation}
to the central slice in both RA and ZA, fixing $\sigma_\mathrm{z}$ and $\mu$ individually for each slice. $\mu$ is simply the central pixel in the slice and $\sigma_\mathrm{z}$ is based on the mean $\sigma$ measured from slices of the well-sampled stacks that include all of the clusters in the simulation volume at $z = 1$ and $2$. Given CHIME's asymmetric primary beam shape, we find $\sigma_\mathrm{RA, z = 1} = 1.62 \pm 0.01$ pix, $\sigma_\mathrm{RA, z = 2} = 2.41 \pm 0.04$ pix, $\sigma_\mathrm{ZA, z = 1} = 2.73 \pm 0.04$ pix, and $\sigma_\mathrm{ZA, z = 2} = 3.08 \pm 0.06$ pix. 
Here $x$ is in pixels and $n$ is the maximum amplitude of the curve. $F_\mathrm{pk, sub}$ is simply the average of $n_\mathrm{RA}$ and $n_\mathrm{ZA}$\footnote{For stacks based on large catalogs (with $N_\mathrm{targets} \gg 1000$), this Gaussian fitting method is comparable to using $F_\mathrm{pk,sub}$ taken directly from the amplitude of the central pixel of the background-subtracted stack. The approach we detail in Section \ref{sec:mass} has the advantage of being more consistent across sample sizes, though it is slightly more computationally intensive due to the fitting.}, which is useful in mitigating some of the effects of noise on the measured peak.

$C$ accounts for the large beam and low resolution characteristic of 21\,cm intensity mapping experiments, and can be expressed generically as 
\begin{equation}
    C(z) = f_\mathrm{ff} \, f_\mathrm{nt}\,(1 + z)^{-2}\,\frac{A_{\mathrm{pix}, z}}{\langle A_\mathrm{HI}\rangle_z} \label{eq:c_gen}
\end{equation}
A more specific expression for $C(z)$ with CHIME is
\begin{equation}
    C(z) = f_\mathrm{ff} \, f_\mathrm{nt}\,(1 + z)^{-2}\,(5.3\times 5.7)\, \frac{r_{\mathrm{amin}, z}^2}{\pi\langle R_\mathrm{HI}\rangle_z^2}
\end{equation}

Here $A_{\mathrm{pix}, z}$ is the physical area of a pixel at $z$, which for CHIME is $(5.3\times 5.7)\, r_{\mathrm{amin}, z}^2$, where $r_{\mathrm{amin}, z}$ is the proper scale of 1 arcmin at a redshift, $z$. $\langle A_\mathrm{HI}\rangle$ corresponds to the average proper area covered by the clusters' \ion{H}{i} , taking $\langle A_\mathrm{HI}\rangle \approx \pi\langle R_\mathrm{HI}\rangle^2$. Here we assume that the radial extent of cluster \ion{H}{i}  is $1.2\,R_{200}$, though there is evidence that it may extend to at least $1.25\, R_{200}$ at $z = 1$ \citep{2016MNRAS.456.3553V}. As we noted earlier, we assume $f_\mathrm{ff} = 0.2$, which is the fraction of flux that survives foreground filtering -- in reality, this value may be somewhat different based on the conditions of the filtering and the 
input catalog; in \citet{2023ApJ...947...16A}, the CHIME stacks on three different eBOSS catalogs have $f_\mathrm{ff}$ between 0.15 and 0.3 with a mean of 0.2.

    We adopt $\langle R_{200} \rangle$ for the galaxy clusters in the TNG300 at each of $z = 1$ and 2 in estimating $C(z)$.  We find that $\langle R_{200} \rangle_{z = 2} = 0.527 \, h^{-1}$ and $\langle R_{200} \rangle_{z = 1} = 0.763 \, h^{-1}$ cMpc respectively in TNG300. $f_\mathrm{nt}$ is the fraction of the stacked flux that is attributable to non-target sources such as field galaxies. We measure this quantity by looking at the relative power in the cluster stacks used throughout this analysis and cluster stacks on cluster-only maps that do not include field galaxies and are constructed based on the individual cluster populations (i.e., mass-selected or radius-selected). In our noiseless CHIME-like TNG300 maps, across cluster populations and redshifts (accounting for the differing pixel size), $f_\mathrm{nt} = 6.4 \pm 0.2$, which we adopt here, noting that this value should be similar to the real one for CHIME's angular resolution and beam size given the accuracy of the TNG300 clustering, but is likely not exact. At $z = 2$, there are $139\times$ more field galaxies than cluster galaxies and those field galaxies contain $56\times$ more mass than the cluster galaxies. At $z =1$ there are $62\times$ as many field galaxies as cluster galaxies, with the field galaxies containing $47\times$ more  \ion{H}{i}  than cluster galaxies by mass. In this context, the small correction with $f_\mathrm{nt} = 6.4$ is further indication that we are actually measuring the  \ion{H}{i}  signal from the clusters on which we are stacking.
    There is negligible redshift evolution in $f_\mathrm{nt}$, though the different cluster populations have slightly different values of $f_\mathrm{nt}$ associated with them, likely due to their differing extents and  \ion{H}{i}  masses -- $f_\mathrm{nt,mass-selected} = 6.9\pm0.5$ and $f_\mathrm{nt,radius-selected} = 5.8\pm0.3$ -- the averaged value of $f_\mathrm{nt} = 6.4$ results in sufficiently accurate mass estimates. This gives us $C_{z = 2} = 18.75$ and $C_{z = 1} = 8.26$. $C$ could potentially be further tuned by assuming a different $R_{200}(z)$ or extent of  \ion{H}{i}  within the halo.
    
For the stacks and expression we use here, these simple conversion factors put us within a factor of $\lesssim 1.25$ of the measured mean mass in the simulation. At $z = 2$, we find $\langle M_\mathrm{stack} \rangle/\langle M_\mathrm{sim} \rangle = $ $1.26\pm0.02$, $1.24\pm0.02$, and $0.86\pm0.03$ for all galaxy clusters, mass-selected galaxy clusters, and radius-selected galaxy clusters respectively. At $z = 1$, where we stack on many more radius-selected clusters in particular: $\langle M_\mathrm{stack} \rangle/\langle M_\mathrm{sim} \rangle = $ $1.000 \pm 0.009$, $0.982 \pm 0.009$, and $0.93 \pm 0.01$ respectively for all galaxy clusters, mass-selected galaxy clusters, and radius-selected galaxy clusters. The presented range includes the effect of noise, taking the uncertainty to be the width of the distribution of mass estimates after 1000 realizations of randomly generated background stacks. The underprediction of the radius-selected galaxy cluster mass at $z = 2$ is driven by our single assumed average $R_{200}$ being somewhat too small for the selection criterion, which is why the radius-selected clusters at $z = 1$ are better matched. As mentioned above, $C(z)$ can be refined by considering a different $\langle R_{200} \rangle$ for a particular redshift, or even more specifically, for the target clusters at each redshift. Similarly, the overprediction of mass in mass-selected clusters (and our full cluster sample given the preponderance of mass-selected clusters) at $z = 2$ is likely the result of that selection criterion being defined at $z = 0$ and then traced back along the Main Progenitor Branch, so that this sample includes protoclusters in addition to already evolved clusters. As a result, the gas properties and distribution of the still-assembling higher $z$ mass-selected clusters may be somewhat different than radius-selected clusters at the same redshift or true clusters at $z = 1$ \citep[e.g.,][]{2023ApJ...945L..28D}. This is consistent with the slight increase in $M_\mathrm{HI}$ in mass-selected clusters from $z = 2$ to $z = 1$ as seen in Fig. \ref{fig:clustermass}.

Though the mass recovery strategy we employ here is simple, it serves as an example of how one might go about deriving a cluster \ion{H}{i} mass from stacked \tcm~data. In the case that the fraction of non-target contamination ($f_\mathrm{nt}$ here) and the instrument's flux amplitude calibration are not known to a sufficient degree to infer  \ion{H}{i}  mass directly, it is still possible to infer the relative mass between tracers and/or redshifts. Using the background-subtracted stacks described above and, if comparing across redshifts, the $A_\mathrm{pix,}z/\langle A_\mathrm{HI}\rangle_z$ correction\footnote{To arrive at this corrective factor, we assume that (i) the distribution and properties of galaxies and clusters in TNG300 are comparable to those in the real universe, (ii) the  \ion{H}{i}  masses in the \textit{Molecular and atomic hydrogen (HI+H2) galaxy contents} catalogs are realistic, (iii) equation (\ref{eq:draine}) is accurate for all source types, and (iv) the CHIME beam is correct (noting that we already use a \textit{Y}-polarized beam in this analysis instead of the actual \textit{X}+\textit{Y}-polarized beam). With this in mind, we urge some caution in extrapolating this factor to other experiments, suggesting instead that this analysis should be repeated for different beams, as in the scenario where assumptions (1)-(3) are true, (4) ensures that the correction is only informative for the CHIME dataset.}, it is straightforward to return the inferred mass ratio between two independently selected cluster samples. For example, one could then compare the evolution of Sunyaev-Zel'dovich-selected clusters \citep[e.g.,][]{1972CoASP...4..173S, 2014AA...571A..29P, 2015ApJS..216...27B, 2021ApJS..253....3H} to the evolution of X-ray-selected clusters using only a pixel size normalization (without assuming $f_\mathrm{nt}$ or $\langle R_\mathrm{HI} \rangle$) even without having a fully accurate flux calibration in place.

\subsection{The extent of  \ion{H}{i}  in $z = 2, \, 1$ TNG300 clusters}

One innate limitation of this analysis is that the mass resolution of the \textit{Molecular and atomic hydrogen (\textsc{HI+H2}) galaxy contents} catalog limits us to looking at the gas content of the most massive galaxies. These massive galaxies generally experience the most efficient quenching in cluster environments \citep[e.g.][]{2018ApJ...866..136F, 2019ApJ...876...40P}, and in that sense, we are adequately exploring what happens to the neutral hydrogen in galaxy clusters. However, it is not clear that we are sensitive to the gaseous intracluster medium (ICM) or the diffuse cluster outskirts in this simulation using the existing \textit{Molecular and atomic hydrogen (\textsc{HI+H2}) galaxy contents} catalogs, since they tie  \ion{H}{i}  mass to individual subhaloes rather than to the larger halo; the ICM is generally thought to be too hot and ionized to contain substantial amounts of neutral hydrogen, but the clusters' outskirts may house cool gas \citep{2012ARAA..50..353K}. 

\begin{figure*}
    \centering
    \includegraphics[width=2\columnwidth]{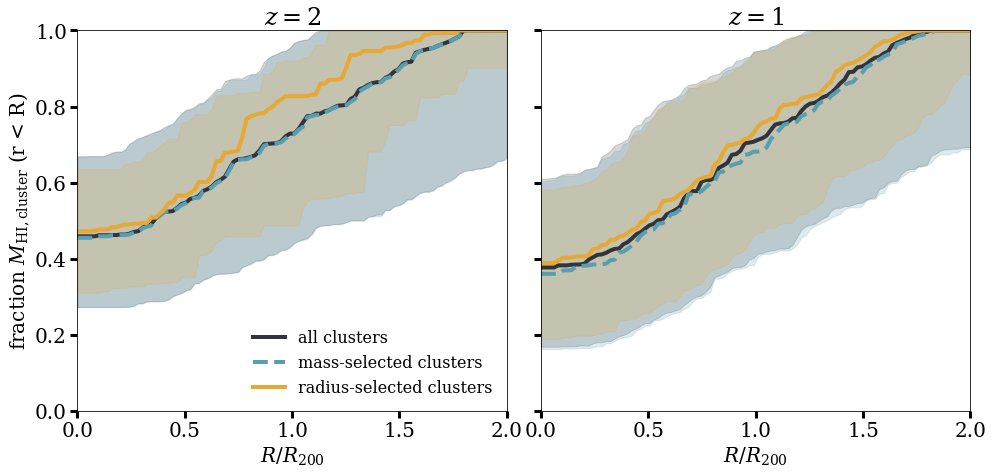}
    \caption{The cumulative normalized  \ion{H}{i}  mass as a function of $R/R_{200}$ for clusters included in our stacks. The lines show the median $M_\mathrm{HI}/M_\mathrm{HI, tot}$ vs. $R/R_{200}$ with the shaded regions corresponding to the 16th and 84th percentile across clusters at that redshift. As the clusters evolve with time, more mass is concentrated in the cluster outskirts with most  \ion{H}{i}  mass within $2\, R_{200}$.
    }
    \label{fig:mass-rad}
\end{figure*}

To better understand where the mass is in the clusters we target in this analysis, we look at the cumulative distributions of  \ion{H}{i}  mass with radius (normalized to $R_{200}$) in Fig. \ref{fig:mass-rad}. It is apparent that $\sim 40$ per cent of the mass in the clusters is outside of $R_{200}$, with $\sim 80$ per cent of the total $M_\mathrm{HI}$ contained in $1.2\, R_{200}$. In the roughly 2.5 Gyr between $z = 2$ and $z = 1$, there is already apparent evolution in the amount of mass concentrated in the centre of the cluster rather than in the outskirts, with lower redshift clusters generally having more neutral hydrogen in their outskirts. This is consistent with observations at $z = 0$, where \textit{most} neutral hydrogen is observed in the cluster outskirts at $R \gtrsim R_{200}$ \citep[e.g.,][]{1985ApJ...292..404G, 2012ApJ...754...84Y}.

Given that the spatial extent of \ion{H}{i} (and the fraction of $M_\mathrm{HI}$ within 1.2\,$R_{200}$) is similar at $z = 2$ and $z = 1$, the direct comparison of the relative  \ion{H}{i}  mass at these two redshifts is feasible. This suggests that CHIME will be able to place observational constraints on the evolution of clusters'  \ion{H}{i}  reservoir between $z = 2$ and $z = 1$, and even a normalized flux ratio will prove diagnostic between redshifts.

\subsection{Minimum sample size for accurate mass recovery}
We have established that it is possible to recover the  \ion{H}{i}  mass when stacking on all of the galaxy clusters in our CHIME-like TNG300 map. In reality, though, there are only a limited number of known galaxy (proto-)clusters at $z = 2$ and $z = 1$ that fall within CHIME's sky coverage. We consider a handful of scenarios, stacking on limited samples of clusters.

To do this, we randomly sample $N$ clusters from the full catalogs without replacement (though we note that the full catalog is based on target locations in the tiled map and includes the same TNG300 clusters in different locations and at different distances), looking explicitly at samples containing 3000, 1000, 300, 100, and 30 clusters. We iterate over the selection and mass estimate 1000 times for each sample size and distinct cluster population (i.e., all clusters, mass-selected clusters, and radius-selected clusters) to place stronger constraints on the accuracy while mitigating bias. Due to this randomization, some stacks may be based on catalogs of less massive, less luminous clusters at large distances, making it possible that small, but thoughtfully selected catalogs will yield more consistent results in reality.

\begin{table*}
	\centering
	\caption{$\langle M_\mathrm{HI,stack}\rangle/\langle M_\mathrm{HI,sim}\rangle$ for stacks on $N$ galaxy clusters. Here \emph{all} denotes that the sample was drawn from a catalog containing all galaxy clusters, \emph{mass-selected} indicates that the sample was drawn from a catalog of mass-selected galaxy clusters, and \emph{radius-selected} indicates that the sample was drawn from a catalog of radius-selected galaxy clusters. The column without a label uses every cluster in the catalog in the stack.}
	\label{tab:samp_acc}
	\begin{tabular}{lcccccc} 
		\hline
		Sample &  & $N = 3000$ & 1000 & 300 & 100 & 30\\
		\hline
		\emph{all}, $z = 2$ & $1.26 \pm 0.02$ & $1.3\pm0.2$ & $1.2\pm0.3$ & $1.2\pm0.5$ & $1.3\pm0.9$ & $1\pm2$\\
            \emph{all}, $z = 1$ & $1.000 \pm 0.009$ & $1.00\pm0.08$ & $1.0\pm0.1$ & $1.0\pm0.3$ & $1.0\pm0.4$ & $1.1\pm0.8$\\
            \emph{mass-selected}, $z = 2$ & $1.24\pm0.02$ & $1.2\pm0.2$ & $1.3\pm0.3$ & $1.3\pm0.5$ & $1.2\pm0.9$ & $1\pm2$\\
            \emph{mass-selected}, $z = 1$ & $0.982 \pm 0.009$ & $0.98\pm0.08$ & $1.0\pm0.1$ & $1.0\pm0.2$ & $1.0\pm0.4$ & $1.0\pm0.8$\\
            \emph{radius-selected}, $z = 2$ & $0.86 \pm 0.03$ & $0.86\pm0.08$ & $0.9\pm0.1$ & $0.9\pm0.3$ & $0.8\pm0.5$ & $0.9\pm0.9$\\
            \emph{radius-selected}, $z = 1$ & $0.93 \pm 0.01$ & $0.93\pm0.06$ & $0.9\pm0.1$ & $0.9\pm0.2$ & $0.9\pm0.4$ & $0.9\pm0.7$\\
		\hline
	\end{tabular}
\end{table*}

Our results are in Table \ref{tab:samp_acc}, and are indicative of the expectation that stacking on more targets does not impact the measured flux, but does affect the noise properties of the stacks. We find that it is possible to recover $\langle M_\mathrm{HI}\rangle$ of the stacked galaxy clusters from a sample at $z = 1$ or $z = 2$ with $\sim$hundreds of targets. This is consistent with the size of current $z \gtrsim 1$ galaxy cluster catalogs \citep[e.g.,][]{2021MNRAS.500..358B, 2021MNRAS.500.1003W, 2021ApJS..253....3H, 2022MNRAS.513.3946W, bulbul2024srgerosita, 2024arXiv240614754K}, though many of these clusters are only assigned a photometric $z$. While we include only a single corrective factor to approximate the impact of foreground filtering in this proof-of-concept analysis, we note that the redshift uncertainties need to be less than or comparable to the velocity dispersion of galaxy clusters ($dz \lesssim 0.005$) to avoid significant attenuation of the signal \citep{2023ApJ...947...16A}, even for projected improvements in foreground filtering. (The loss of flux contributed by high-velocity \ion{H}{i} associated with narrower band stacks should be secondary to the mitigation of signal loss due to filtering, though thorough investigation of this point is beyond the scope of this paper.)
This constraint means that a primary limitation will be building sufficiently large catalogs of clusters with spectroscopic redshift measurements. There is significant promise that the next generation of instruments that are beginning to come online now will substantially expand intermediate-to-high redshift cluster catalogs \citep{2019BAAS...51c.279M, 2020AA...642A..17Z, 2023arXiv231204253C} and offer opportunities for spectroscopic confirmation of existing $z \gtrsim 1$ galaxy cluster candidates. The timing could not be better for CHIME to be equipped to carry out these analyses.

Additionally, we tested a scenario with a CHIME-like signal level at $z = 1$ where we could do a direct comparison and found that it did not significantly impact the result that a mass could be recovered for samples with $N \gtrsim 100$ clusters. Instead, this somewhat lower flux case had mildly increased errors, consistent with the slightly lower SNR, that placed the lower bound on $N$ closer to 300, but which were not large enough to make mass recovery impossible in the $N = 100$ case. Spectroscopic cluster catalogs at this redshift are expected to include hundreds-to-thousands of clusters, so this difference in required sample size is essentially negligible.

\section{Summary \& Conclusions} \label{sec:discuss}

In this work, we discuss the possibility of cosmology experiments like CHIME providing an avenue for us to measure the integrated  \ion{H}{i}  content of clusters at $z = 2 - 1$. 

To summarize our results:
\begin{itemize}
    \item Stacks on cluster galaxies show substantially more  \ion{H}{i}  flux than stacks on field galaxies (Fig. \ref{fig:stackex} and Fig. \ref{fig:galstacks}). This is likely due to the large beam covering the entire cluster, making the stack sensitive to the cluster's neutral hydrogen content.
    \item Radius-selected clusters contain, on average, more neutral hydrogen mass than mass-selected clusters at both $z = 2$ and $z = 1$ (see Fig. \ref{fig:clusterstacks} and Fig. \ref{fig:clustermass}). Because they are selected for their properties at $z = 2$ or 1 rather than their evolved properties at $z = 0$, radius-selected clusters are intrinsically larger and more massive than mass-selected clusters at the same redshift.
    \item We recover from TNG300 the evolution of galaxy clusters to have more  \ion{H}{i}  in their outskirts from $z = 2$ to $z = 1$ (Fig. \ref{fig:mass-rad}). This is consistent with low redshift observations of galaxy clusters, for which most neutral hydrogen mass is measured outside at $R \gtrsim R_{200}$.
    \item It is possible to recover a mean cluster $M_\mathrm{HI}$ from stacks as long as there are $N \gtrsim 100$ clusters in the sample (Table \ref{tab:samp_acc}). This is promising for understanding the properties of high redshift galaxy clusters given current catalogs and the forecast for expanded samples anticipated from next-generation instruments. Similarly, this approach can be used to infer the  \ion{H}{i}  mass ratio between redshifts and/or cluster selection criteria even without a robust instrument flux calibration. Comparison between redshifts to derive a relative abundance requires a simple normalization for the change in pixel proper scale, while comparison between cluster catalogs at the same redshift can be done directly without correction. We note, though, that depending on the exact protocol for converting between measured flux and \ion{H}{i} content, assumptions like those about the neutral hydrogen volume filling factor or the spatial extent of the typical galaxy clusters may introduce additional redshift dependence that complicates mass ratio recovery across redshifts.
\end{itemize}

The primary limitations in measuring the evolution of neutral hydrogen in intermediate-to-high redshift galaxy clusters and protoclusters are the instrument sensitivity and resolution compared to the scale and \ion{H}{i} mass of the target objects \citep{2004MNRAS.355.1339B, 2005AJ....130.2598G}. Since CHIME is able to study the average, stacked \ion{H}{i} content of clusters that are not resolved by the beam, these results may also have applications for other instruments of comparable resolution and sky coverage. For instance, the CHIME synthesized beam and the GMRT primary beam have similar angular extents in the same redshift range \citep[e.g.][]{2022arXiv220515334C}, and the Hydrogen Intensity and Real-time Analysis eXperiment \citep[HIRAX,][]{2016SPIE.9906E..5XN, 2022JATIS...8a1019C} will offer a direct complement to CHIME, mapping the \ion{H}{i} signal in the Southern sky with the same frequency coverage and spatial resolution. As catalogs of galaxy clusters at $z \ge 1$~grow, \tcm experiments offer a promising avenue to study the evolving \ion{H}{i} reservoir in clusters and proto-clusters. Such studies should also be able to provide constraints on the character and strength of feedback mechanisms that are active in cluster environments as a function of redshift.

\section*{Acknowledgements}
The authors thank Frank van den Bosch and H\'{e}ctor Arce for a fruitful discussion that helped this paper take its initial shape and thank the anonymous referee for comments and suggestions that improved the manuscript. AP thanks Seth Siegel, Simon Foreman, and Richard Shaw, along with the rest of the Canadian Hydrogen Intensity Mapping Experiment Collaboration, for help generating CHIME-like beams, assessing simulation suitability, and useful comments that enhanced this work.

\section*{Data Availability}

Data used in this paper are publicly available as part of TNG300 simulation \citep[][]{2018MNRAS.475..648P, 2018MNRAS.475..676S, 2018MNRAS.475..624N, 2018MNRAS.477.1206N, 2018MNRAS.480.5113M} from the IllustrisTNG project\footnote{\href{https://www.tng-project.org/data/downloads/TNG300-1/}{https://www.tng-project.org/data/downloads/TNG300-1/}} \citep[][]{2019ComAC...6....2N}. Additionally, our beams, catalogs, tiled intensity maps, and analysis code are available on GitHub\footnote{\href{https://github.com/avapolzin/CHIMExIllustrisTNG}{https://github.com/avapolzin/CHIMExIllustrisTNG}}.

\bibliographystyle{mnras}
\bibliography{references}

\begin{thebibliography}{}
\makeatletter
\relax
\def\mn@urlcharsother{\let\do\@makeother \do\$\do\&\do\#\do\^\do\_\do\%\do\~}
\def\mn@doi{\begingroup\mn@urlcharsother \@ifnextchar [ {\mn@doi@}
  {\mn@doi@[]}}
\def\mn@doi@[#1]#2{\def\@tempa{#1}\ifx\@tempa\@empty \href
  {http://dx.doi.org/#2} {doi:#2}\else \href {http://dx.doi.org/#2} {#1}\fi
  \endgroup}
\def\mn@eprint#1#2{\mn@eprint@#1:#2::\@nil}
\def\mn@eprint@arXiv#1{\href {http://arxiv.org/abs/#1} {{\tt arXiv:#1}}}
\def\mn@eprint@dblp#1{\href {http://dblp.uni-trier.de/rec/bibtex/#1.xml}
  {dblp:#1}}
\def\mn@eprint@#1:#2:#3:#4\@nil{\def\@tempa {#1}\def\@tempb {#2}\def\@tempc
  {#3}\ifx \@tempc \@empty \let \@tempc \@tempb \let \@tempb \@tempa \fi \ifx
  \@tempb \@empty \def\@tempb {arXiv}\fi \@ifundefined
  {mn@eprint@\@tempb}{\@tempb:\@tempc}{\expandafter \expandafter \csname
  mn@eprint@\@tempb\endcsname \expandafter{\@tempc}}}

\bibitem[\protect\citeauthoryear{{Abazajian} et~al.,}{{Abazajian}
  et~al.}{2016}]{2016arXiv161002743A}
{Abazajian} K.~N.,  et~al., 2016, \mn@doi [arXiv e-prints]
  {10.48550/arXiv.1610.02743}, \href
  {https://ui.adsabs.harvard.edu/abs/2016arXiv161002743A} {p. arXiv:1610.02743}

\bibitem[\protect\citeauthoryear{{Ade} et~al.,}{{Ade}
  et~al.}{2019}]{2019JCAP...02..056A}
{Ade} P.,  et~al., 2019, \mn@doi [\jcap] {10.1088/1475-7516/2019/02/056}, \href
  {https://ui.adsabs.harvard.edu/abs/2019JCAP...02..056A} {2019, 056}

\bibitem[\protect\citeauthoryear{{Ai}, {Zhu}  \& {Fu}}{{Ai}
  et~al.}{2017}]{2017RAA....17..101A}
{Ai} M.,  {Zhu} M.,   {Fu} J.,  2017, \mn@doi [Research in Astronomy and
  Astrophysics] {10.1088/1674-4527/17/10/101}, \href
  {https://ui.adsabs.harvard.edu/abs/2017RAA....17..101A} {17, 101}

\bibitem[\protect\citeauthoryear{{Balogh} et~al.,}{{Balogh}
  et~al.}{2021}]{2021MNRAS.500..358B}
{Balogh} M.~L.,  et~al., 2021, \mn@doi [\mnras] {10.1093/mnras/staa3008}, \href
  {https://ui.adsabs.harvard.edu/abs/2021MNRAS.500..358B} {500, 358}

\bibitem[\protect\citeauthoryear{{Battye}, {Davies}  \& {Weller}}{{Battye}
  et~al.}{2004}]{2004MNRAS.355.1339B}
{Battye} R.~A.,  {Davies} R.~D.,   {Weller} J.,  2004, \mn@doi [\mnras]
  {10.1111/j.1365-2966.2004.08416.x}, \href
  {https://ui.adsabs.harvard.edu/abs/2004MNRAS.355.1339B} {355, 1339}

\bibitem[\protect\citeauthoryear{{Bleem} et~al.,}{{Bleem}
  et~al.}{2015}]{2015ApJS..216...27B}
{Bleem} L.~E.,  et~al., 2015, \mn@doi [\apjs] {10.1088/0067-0049/216/2/27},
  \href {https://ui.adsabs.harvard.edu/abs/2015ApJS..216...27B} {216, 27}

\bibitem[\protect\citeauthoryear{Bulbul et~al.,}{Bulbul
  et~al.}{2024}]{bulbul2024srgerosita}
Bulbul E.,  et~al., 2024, The SRG/eROSITA All-Sky Survey: The first catalog of
  galaxy clusters and groups in the Western Galactic Hemisphere (\mn@eprint
  {arXiv} {2402.08452})

\bibitem[\protect\citeauthoryear{{CHIME Collaboration} et~al.,}{{CHIME
  Collaboration} et~al.}{2022a}]{2022ApJS..261...29C}
{CHIME Collaboration} et~al., 2022a, \mn@doi [\apjs]
  {10.3847/1538-4365/ac6fd9}, \href
  {https://ui.adsabs.harvard.edu/abs/2022ApJS..261...29C} {261, 29}

\bibitem[\protect\citeauthoryear{{CHIME Collaboration} et~al.,}{{CHIME
  Collaboration} et~al.}{2022b}]{2022ApJ...932..100A}
{CHIME Collaboration} et~al., 2022b, \mn@doi [\apj] {10.3847/1538-4357/ac6b9f},
  \href {https://ui.adsabs.harvard.edu/abs/2022ApJ...932..100A} {932, 100}

\bibitem[\protect\citeauthoryear{{CHIME Collaboration} et~al.,}{{CHIME
  Collaboration} et~al.}{2023}]{2023ApJ...947...16A}
{CHIME Collaboration} et~al., 2023, \mn@doi [\apj] {10.3847/1538-4357/acb13f},
  \href {https://ui.adsabs.harvard.edu/abs/2023ApJ...947...16A} {947, 16}

\bibitem[\protect\citeauthoryear{{CHIME Collaboration} et~al.,}{{CHIME
  Collaboration} et~al.}{2024}]{2024ApJ...963...23A}
{CHIME Collaboration} et~al., 2024, \mn@doi [\apj] {10.3847/1538-4357/ad0f1d},
  \href {https://ui.adsabs.harvard.edu/abs/2024ApJ...963...23A} {963, 23}

\bibitem[\protect\citeauthoryear{{Cerardi}, {Pierre}, {Valageas}, {Garrel}  \&
  {Pacaud}}{{Cerardi} et~al.}{2023}]{2023arXiv231204253C}
{Cerardi} N.,  {Pierre} M.,  {Valageas} P.,  {Garrel} C.,   {Pacaud} F.,  2023,
  \mn@doi [arXiv e-prints] {10.48550/arXiv.2312.04253}, \href
  {https://ui.adsabs.harvard.edu/abs/2023arXiv231204253C} {p. arXiv:2312.04253}

\bibitem[\protect\citeauthoryear{{Chang}, {Pen}, {Peterson}  \&
  {McDonald}}{{Chang} et~al.}{2008}]{2008PhRvL.100i1303C}
{Chang} T.-C.,  {Pen} U.-L.,  {Peterson} J.~B.,   {McDonald} P.,  2008, \mn@doi
  [\prl] {10.1103/PhysRevLett.100.091303}, \href
  {https://ui.adsabs.harvard.edu/abs/2008PhRvL.100i1303C} {100, 091303}

\bibitem[\protect\citeauthoryear{{Chowdhury}, {Kanekar}  \&
  {Chengalur}}{{Chowdhury} et~al.}{2022}]{2022arXiv220515334C}
{Chowdhury} A.,  {Kanekar} N.,   {Chengalur} J.~N.,  2022, arXiv e-prints,
  \href {https://ui.adsabs.harvard.edu/abs/2022arXiv220515334C} {p.
  arXiv:2205.15334}

\bibitem[\protect\citeauthoryear{{Crichton} et~al.,}{{Crichton}
  et~al.}{2022}]{2022JATIS...8a1019C}
{Crichton} D.,  et~al., 2022, \mn@doi [Journal of Astronomical Telescopes,
  Instruments, and Systems] {10.1117/1.JATIS.8.1.011019}, \href
  {https://ui.adsabs.harvard.edu/abs/2022JATIS...8a1019C} {8, 011019}

\bibitem[\protect\citeauthoryear{{D'Aloisio}, {Furlanetto}  \&
  {Natarajan}}{{D'Aloisio} et~al.}{2009}]{2009MNRAS.394.1469D}
{D'Aloisio} A.,  {Furlanetto} S.~R.,   {Natarajan} P.,  2009, \mn@doi [\mnras]
  {10.1111/j.1365-2966.2009.14400.x}, \href
  {https://ui.adsabs.harvard.edu/abs/2009MNRAS.394.1469D} {394, 1469}

\bibitem[\protect\citeauthoryear{{Dawson} et~al.,}{{Dawson}
  et~al.}{2016}]{2016AJ....151...44D}
{Dawson} K.~S.,  et~al., 2016, \mn@doi [\aj] {10.3847/0004-6256/151/2/44},
  \href {https://ui.adsabs.harvard.edu/abs/2016AJ....151...44D} {151, 44}

\bibitem[\protect\citeauthoryear{{Diemer} et~al.,}{{Diemer}
  et~al.}{2018}]{2018ApJS..238...33D}
{Diemer} B.,  et~al., 2018, \mn@doi [\apjs] {10.3847/1538-4365/aae387}, \href
  {https://ui.adsabs.harvard.edu/abs/2018ApJS..238...33D} {238, 33}

\bibitem[\protect\citeauthoryear{{Diemer} et~al.,}{{Diemer}
  et~al.}{2019}]{2019MNRAS.487.1529D}
{Diemer} B.,  et~al., 2019, \mn@doi [\mnras] {10.1093/mnras/stz1323}, \href
  {https://ui.adsabs.harvard.edu/abs/2019MNRAS.487.1529D} {487, 1529}

\bibitem[\protect\citeauthoryear{{Dong}, {Lee}, {Ata}, {Horowitz}  \&
  {Momose}}{{Dong} et~al.}{2023}]{2023ApJ...945L..28D}
{Dong} C.,  {Lee} K.-G.,  {Ata} M.,  {Horowitz} B.,   {Momose} R.,  2023,
  \mn@doi [\apjl] {10.3847/2041-8213/acba89}, \href
  {https://ui.adsabs.harvard.edu/abs/2023ApJ...945L..28D} {945, L28}

\bibitem[\protect\citeauthoryear{{Draine}}{{Draine}}{2011}]{2011piim.book.....D}
{Draine} B.~T.,  2011, {Physics of the Interstellar and Intergalactic Medium}

\bibitem[\protect\citeauthoryear{{Faber} et~al.,}{{Faber}
  et~al.}{2007}]{2007ApJ...665..265F}
{Faber} S.~M.,  et~al., 2007, \mn@doi [\apj] {10.1086/519294}, \href
  {https://ui.adsabs.harvard.edu/abs/2007ApJ...665..265F} {665, 265}

\bibitem[\protect\citeauthoryear{{Foltz} et~al.,}{{Foltz}
  et~al.}{2018}]{2018ApJ...866..136F}
{Foltz} R.,  et~al., 2018, \mn@doi [\apj] {10.3847/1538-4357/aad80d}, \href
  {https://ui.adsabs.harvard.edu/abs/2018ApJ...866..136F} {866, 136}

\bibitem[\protect\citeauthoryear{{Francis}, {Wilson}  \& {Woodgate}}{{Francis}
  et~al.}{2001}]{2001PASA...18...64F}
{Francis} P.~J.,  {Wilson} G.~M.,   {Woodgate} B.~E.,  2001, \mn@doi [\pasa]
  {10.1071/AS01005}, \href
  {https://ui.adsabs.harvard.edu/abs/2001PASA...18...64F} {18, 64}

\bibitem[\protect\citeauthoryear{{Giovanelli} \& {Haynes}}{{Giovanelli} \&
  {Haynes}}{1985}]{1985ApJ...292..404G}
{Giovanelli} R.,  {Haynes} M.~P.,  1985, \mn@doi [\apj] {10.1086/163170}, \href
  {https://ui.adsabs.harvard.edu/abs/1985ApJ...292..404G} {292, 404}

\bibitem[\protect\citeauthoryear{{Giovanelli} et~al.,}{{Giovanelli}
  et~al.}{2005}]{2005AJ....130.2598G}
{Giovanelli} R.,  et~al., 2005, \mn@doi [\aj] {10.1086/497431}, \href
  {https://ui.adsabs.harvard.edu/abs/2005AJ....130.2598G} {130, 2598}

\bibitem[\protect\citeauthoryear{{Gnedin} \& {Kravtsov}}{{Gnedin} \&
  {Kravtsov}}{2011}]{2011ApJ...728...88G}
{Gnedin} N.~Y.,  {Kravtsov} A.~V.,  2011, \mn@doi [\apj]
  {10.1088/0004-637X/728/2/88}, \href
  {https://ui.adsabs.harvard.edu/abs/2011ApJ...728...88G} {728, 88}

\bibitem[\protect\citeauthoryear{{Guo} et~al.,}{{Guo}
  et~al.}{2019}]{2019ApJ...871..147G}
{Guo} H.,  et~al., 2019, \mn@doi [\apj] {10.3847/1538-4357/aaf9ad}, \href
  {https://ui.adsabs.harvard.edu/abs/2019ApJ...871..147G} {871, 147}

\bibitem[\protect\citeauthoryear{{Hilton} et~al.,}{{Hilton}
  et~al.}{2021}]{2021ApJS..253....3H}
{Hilton} M.,  et~al., 2021, \mn@doi [\apjs] {10.3847/1538-4365/abd023}, \href
  {https://ui.adsabs.harvard.edu/abs/2021ApJS..253....3H} {253, 3}

\bibitem[\protect\citeauthoryear{{Jiang} et~al.,}{{Jiang}
  et~al.}{2019}]{2019SCPMA..6259502J}
{Jiang} P.,  et~al., 2019, \mn@doi [Science China Physics, Mechanics, and
  Astronomy] {10.1007/s11433-018-9376-1}, \href
  {https://ui.adsabs.harvard.edu/abs/2019SCPMA..6259502J} {62, 959502}

\bibitem[\protect\citeauthoryear{{Jiang} et~al.,}{{Jiang}
  et~al.}{2020}]{2020RAA....20...64J}
{Jiang} P.,  et~al., 2020, \mn@doi [Research in Astronomy and Astrophysics]
  {10.1088/1674-4527/20/5/64}, \href
  {https://ui.adsabs.harvard.edu/abs/2020RAA....20...64J} {20, 064}

\bibitem[\protect\citeauthoryear{{Klein}, {Mohr}  \& {Davies}}{{Klein}
  et~al.}{2024}]{2024arXiv240614754K}
{Klein} M.,  {Mohr} J.~J.,   {Davies} C.~T.,  2024, arXiv e-prints, \href
  {https://ui.adsabs.harvard.edu/abs/2024arXiv240614754K} {p. arXiv:2406.14754}

\bibitem[\protect\citeauthoryear{{Kravtsov} \& {Borgani}}{{Kravtsov} \&
  {Borgani}}{2012}]{2012ARAA..50..353K}
{Kravtsov} A.~V.,  {Borgani} S.,  2012, \mn@doi [\araa]
  {10.1146/annurev-astro-081811-125502}, \href
  {https://ui.adsabs.harvard.edu/abs/2012ARA&A..50..353K} {50, 353}

\bibitem[\protect\citeauthoryear{{Madau} \& {Dickinson}}{{Madau} \&
  {Dickinson}}{2014}]{2014ARAA..52..415M}
{Madau} P.,  {Dickinson} M.,  2014, \mn@doi [\araa]
  {10.1146/annurev-astro-081811-125615}, \href
  {https://ui.adsabs.harvard.edu/abs/2014ARA&A..52..415M} {52, 415}

\bibitem[\protect\citeauthoryear{{Mantz} et~al.,}{{Mantz}
  et~al.}{2019}]{2019BAAS...51c.279M}
{Mantz} A.,  et~al., 2019, \mn@doi [\baas] {10.48550/arXiv.1903.05606}, \href
  {https://ui.adsabs.harvard.edu/abs/2019BAAS...51c.279M} {51, 279}

\bibitem[\protect\citeauthoryear{{Marinacci} et~al.,}{{Marinacci}
  et~al.}{2018}]{2018MNRAS.480.5113M}
{Marinacci} F.,  et~al., 2018, \mn@doi [\mnras] {10.1093/mnras/sty2206}, \href
  {https://ui.adsabs.harvard.edu/abs/2018MNRAS.480.5113M} {480, 5113}

\bibitem[\protect\citeauthoryear{{Naiman} et~al.,}{{Naiman}
  et~al.}{2018}]{2018MNRAS.477.1206N}
{Naiman} J.~P.,  et~al., 2018, \mn@doi [\mnras] {10.1093/mnras/sty618}, \href
  {https://ui.adsabs.harvard.edu/abs/2018MNRAS.477.1206N} {477, 1206}

\bibitem[\protect\citeauthoryear{{Nelson} et~al.,}{{Nelson}
  et~al.}{2018}]{2018MNRAS.475..624N}
{Nelson} D.,  et~al., 2018, \mn@doi [\mnras] {10.1093/mnras/stx3040}, \href
  {https://ui.adsabs.harvard.edu/abs/2018MNRAS.475..624N} {475, 624}

\bibitem[\protect\citeauthoryear{{Nelson} et~al.,}{{Nelson}
  et~al.}{2019}]{2019ComAC...6....2N}
{Nelson} D.,  et~al., 2019, \mn@doi [Computational Astrophysics and Cosmology]
  {10.1186/s40668-019-0028-x}, \href
  {https://ui.adsabs.harvard.edu/abs/2019ComAC...6....2N} {6, 2}

\bibitem[\protect\citeauthoryear{{Newburgh} et~al.,}{{Newburgh}
  et~al.}{2014}]{2014SPIE.9145E..4VN}
{Newburgh} L.~B.,  et~al., 2014, in {Stepp} L.~M.,  {Gilmozzi} R.,   {Hall}
  H.~J.,  eds,  Society of Photo-Optical Instrumentation Engineers (SPIE)
  Conference Series Vol. 9145, Ground-based and Airborne Telescopes V. p.
  91454V (\mn@eprint {arXiv} {1406.2267}), \mn@doi{10.1117/12.2056962}

\bibitem[\protect\citeauthoryear{{Newburgh} et~al.,}{{Newburgh}
  et~al.}{2016}]{2016SPIE.9906E..5XN}
{Newburgh} L.~B.,  et~al., 2016, in {Hall} H.~J.,  {Gilmozzi} R.,   {Marshall}
  H.~K.,  eds,  Society of Photo-Optical Instrumentation Engineers (SPIE)
  Conference Series Vol. 9906, Ground-based and Airborne Telescopes VI. p.
  99065X (\mn@eprint {arXiv} {1607.02059}), \mn@doi{10.1117/12.2234286}

\bibitem[\protect\citeauthoryear{{Peterson} et~al.,}{{Peterson}
  et~al.}{2009}]{2009astro2010S.234P}
{Peterson} J.~B.,  et~al., 2009, in astro2010: The Astronomy and Astrophysics
  Decadal Survey. p.~234 (\mn@eprint {arXiv} {0902.3091}),
  \mn@doi{10.48550/arXiv.0902.3091}

\bibitem[\protect\citeauthoryear{{Pillepich} et~al.,}{{Pillepich}
  et~al.}{2018}]{2018MNRAS.475..648P}
{Pillepich} A.,  et~al., 2018, \mn@doi [\mnras] {10.1093/mnras/stx3112}, \href
  {https://ui.adsabs.harvard.edu/abs/2018MNRAS.475..648P} {475, 648}

\bibitem[\protect\citeauthoryear{{Pintos-Castro}, {Yee}, {Muzzin}, {Old}  \&
  {Wilson}}{{Pintos-Castro} et~al.}{2019}]{2019ApJ...876...40P}
{Pintos-Castro} I.,  {Yee} H.~K.~C.,  {Muzzin} A.,  {Old} L.,   {Wilson} G.,
  2019, \mn@doi [\apj] {10.3847/1538-4357/ab14ee}, \href
  {https://ui.adsabs.harvard.edu/abs/2019ApJ...876...40P} {876, 40}

\bibitem[\protect\citeauthoryear{{Planck Collaboration} et~al.,}{{Planck
  Collaboration} et~al.}{2014}]{2014AA...571A..29P}
{Planck Collaboration} et~al., 2014, \mn@doi [\aap]
  {10.1051/0004-6361/201321523}, \href
  {https://ui.adsabs.harvard.edu/abs/2014A&A...571A..29P} {571, A29}

\bibitem[\protect\citeauthoryear{{Planck Collaboration} et~al.,}{{Planck
  Collaboration} et~al.}{2016}]{2016AA...594A..13P}
{Planck Collaboration} et~al., 2016, \mn@doi [\aap]
  {10.1051/0004-6361/201525830}, \href
  {https://ui.adsabs.harvard.edu/abs/2016A&A...594A..13P} {594, A13}

\bibitem[\protect\citeauthoryear{{Qian}, {Yao}, {Sun}, {Xu}, {Pan}  \&
  {Jiang}}{{Qian} et~al.}{2020}]{2020Innov...100053Q}
{Qian} L.,  {Yao} R.,  {Sun} J.,  {Xu} J.,  {Pan} Z.,   {Jiang} P.,  2020,
  \mn@doi [The Innovation] {10.1016/j.xinn.2020.100053}, \href
  {https://ui.adsabs.harvard.edu/abs/2020Innov...100053Q} {1, 100053}

\bibitem[\protect\citeauthoryear{{Shaw}, {Sigurdson}, {Sitwell}, {Stebbins}  \&
  {Pen}}{{Shaw} et~al.}{2015}]{2015PhRvD..91h3514S}
{Shaw} J.~R.,  {Sigurdson} K.,  {Sitwell} M.,  {Stebbins} A.,   {Pen} U.-L.,
  2015, \mn@doi [\prd] {10.1103/PhysRevD.91.083514}, \href
  {https://ui.adsabs.harvard.edu/abs/2015PhRvD..91h3514S} {91, 083514}

\bibitem[\protect\citeauthoryear{{Springel} et~al.,}{{Springel}
  et~al.}{2018}]{2018MNRAS.475..676S}
{Springel} V.,  et~al., 2018, \mn@doi [\mnras] {10.1093/mnras/stx3304}, \href
  {https://ui.adsabs.harvard.edu/abs/2018MNRAS.475..676S} {475, 676}

\bibitem[\protect\citeauthoryear{{Sunyaev} \& {Zeldovich}}{{Sunyaev} \&
  {Zeldovich}}{1972}]{1972CoASP...4..173S}
{Sunyaev} R.~A.,  {Zeldovich} Y.~B.,  1972, Comments on Astrophysics and Space
  Physics, \href {https://ui.adsabs.harvard.edu/abs/1972CoASP...4..173S} {4,
  173}

\bibitem[\protect\citeauthoryear{{Tremmel} et~al.,}{{Tremmel}
  et~al.}{2019}]{2019MNRAS.483.3336T}
{Tremmel} M.,  et~al., 2019, \mn@doi [\mnras] {10.1093/mnras/sty3336}, \href
  {https://ui.adsabs.harvard.edu/abs/2019MNRAS.483.3336T} {483, 3336}

\bibitem[\protect\citeauthoryear{{Villaescusa-Navarro}
  et~al.,}{{Villaescusa-Navarro} et~al.}{2016}]{2016MNRAS.456.3553V}
{Villaescusa-Navarro} F.,  et~al., 2016, \mn@doi [\mnras]
  {10.1093/mnras/stv2904}, \href
  {https://ui.adsabs.harvard.edu/abs/2016MNRAS.456.3553V} {456, 3553}

\bibitem[\protect\citeauthoryear{{Wen} \& {Han}}{{Wen} \&
  {Han}}{2021}]{2021MNRAS.500.1003W}
{Wen} Z.~L.,  {Han} J.~L.,  2021, \mn@doi [\mnras] {10.1093/mnras/staa3308},
  \href {https://ui.adsabs.harvard.edu/abs/2021MNRAS.500.1003W} {500, 1003}

\bibitem[\protect\citeauthoryear{{Wen} \& {Han}}{{Wen} \&
  {Han}}{2022}]{2022MNRAS.513.3946W}
{Wen} Z.~L.,  {Han} J.~L.,  2022, \mn@doi [\mnras] {10.1093/mnras/stac1149},
  \href {https://ui.adsabs.harvard.edu/abs/2022MNRAS.513.3946W} {513, 3946}

\bibitem[\protect\citeauthoryear{{Wild}}{{Wild}}{1952}]{1952ApJ...115..206W}
{Wild} J.~P.,  1952, \mn@doi [\apj] {10.1086/145533}, \href
  {https://ui.adsabs.harvard.edu/abs/1952ApJ...115..206W} {115, 206}

\bibitem[\protect\citeauthoryear{{Wyithe} \& {Loeb}}{{Wyithe} \&
  {Loeb}}{2008}]{2008MNRAS.383..606W}
{Wyithe} J. S.~B.,  {Loeb} A.,  2008, \mn@doi [\mnras]
  {10.1111/j.1365-2966.2007.12568.x}, \href
  {https://ui.adsabs.harvard.edu/abs/2008MNRAS.383..606W} {383, 606}

\bibitem[\protect\citeauthoryear{{Yoon}, {Putman}, {Thom}, {Chen}  \&
  {Bryan}}{{Yoon} et~al.}{2012}]{2012ApJ...754...84Y}
{Yoon} J.~H.,  {Putman} M.~E.,  {Thom} C.,  {Chen} H.-W.,   {Bryan} G.~L.,
  2012, \mn@doi [\apj] {10.1088/0004-637X/754/2/84}, \href
  {https://ui.adsabs.harvard.edu/abs/2012ApJ...754...84Y} {754, 84}

\bibitem[\protect\citeauthoryear{{Zhang}, {Ramos-Ceja}, {Pacaud}  \&
  {Reiprich}}{{Zhang} et~al.}{2020}]{2020AA...642A..17Z}
{Zhang} C.,  {Ramos-Ceja} M.~E.,  {Pacaud} F.,   {Reiprich} T.~H.,  2020,
  \mn@doi [\aap] {10.1051/0004-6361/201937329}, \href
  {https://ui.adsabs.harvard.edu/abs/2020A&A...642A..17Z} {642, A17}

\makeatother
\end{thebibliography}

\appendix

\section{Making a tiled map from the simulation volume}
\label{app:makemap}

The volume is projected along a single axis, with the galaxy's luminosity distance defined by $d_{L}(z)$~plus its location along the collapsed, projected axis of the volume. We iterate through different projections of the simulation volume in order to construct a larger map that better replicates the angular coverage of real CHIME data.

\begin{figure*}
	\includegraphics[width=1.5\columnwidth]{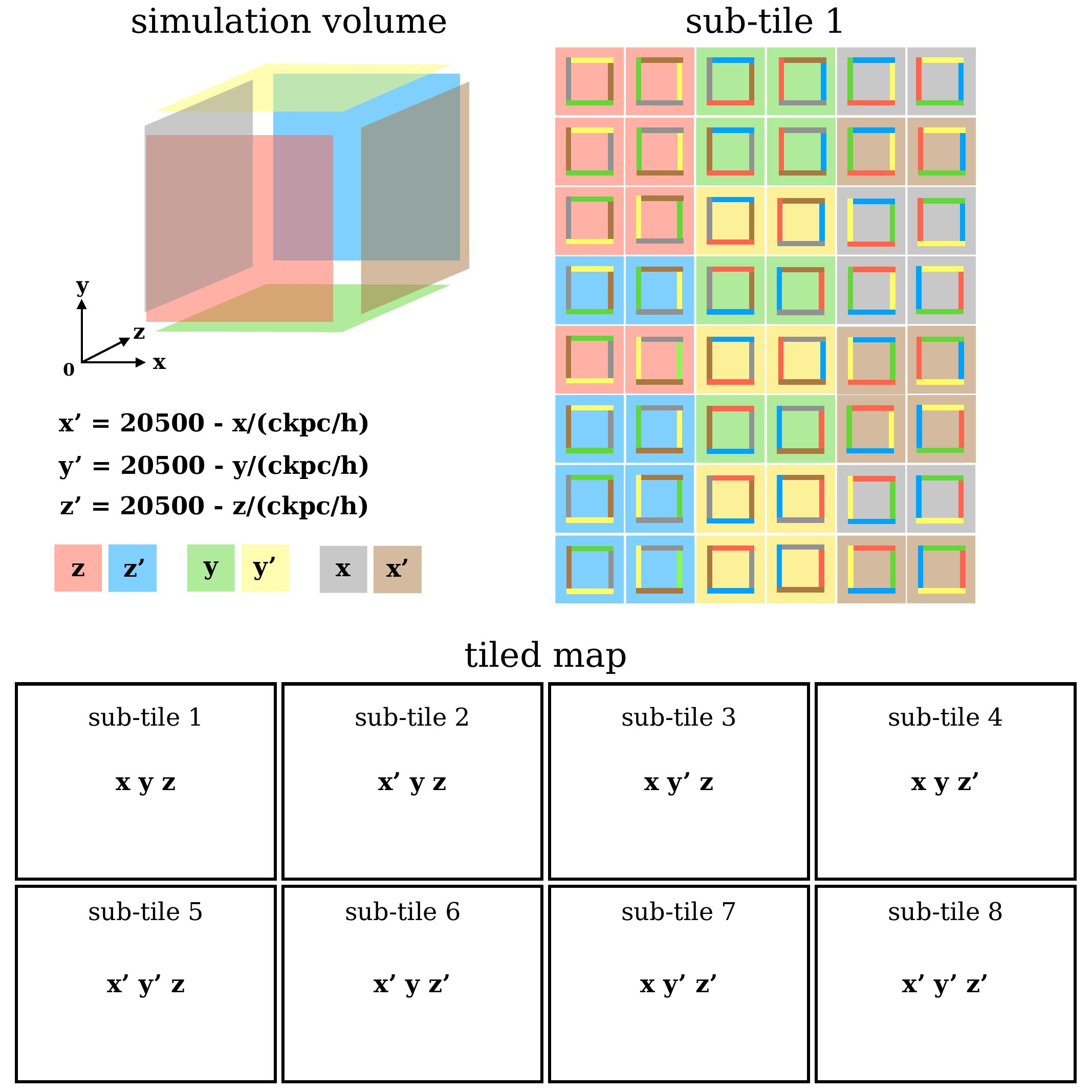}
    \caption{The various permutations of the simulation volume that make up our individual sub-tiles and, in turn, the full tiled map that we use to simulate CHIME observations. The \emph{simulation volume} color codes the different faces of the volume, which are perpendicular to the projection in each case. The face color indicates the side from which $d_\mathrm{L}$ is computed for each target (e.g., for the red face $d_\mathrm{L} = d_\mathrm{L}(\mathrm{redshift}) + z$, where $z$ is the distance along that axis, while for the blue face $d_\mathrm{L} = d_\mathrm{L}(\mathrm{redshift}) + [20500 - z/(\mathrm{ckpc/h})]$). \emph{Sub-tile 1} shows the way different flipped and rotated projections are combined to generate the sub-tiles (using the top left sub-tile as an example). The open squares indicate the orientation of the projection using the face's edges. The \emph{tiled map} demonstrates how the different sub-tiles are ultimately combined, with x' indicating that the x-axis has been flipped for all of the projections in the sub-tile (and so on for y' and z'). In the case of sub-tile 2 then, the red square second from the top and from the left suggests that the simulation volume is projected along the z-axis and included in the sub-tile as x vs. y.}
    \label{fig:makemap}
\end{figure*}

We generate sub-tiles from all possible projections of the simulation volume from flipping and rotating projected maps, and combine together eight such sub-tiles into the tiled map, which is later convolved with the CHIME beam and used in our analysis. The eight sub-tiles are generated in the same way, but each of the sub-tiles has a different subset of the axes flipped for all permutations. We show this in Fig. \ref{fig:makemap}.

While this does lead to some repeated projections in the full tiled map, any repetition is mitigated by the neighboring projections being similarly affected generating differences on larger scales which will be affected by beam-convolution and, ultimately, the addition of noise.

\section{Mock catalog creation}
\label{appA}
As so much of this analysis rests on the catalogs of (mass- or radius-selected) clusters that we use for stacking, we will go into this catalog creation in greater detail for those interested. (The salient points for the analysis itself are included in the main body of the text and all of code used to generate our catalogs is publicly available.)

\subsection{Mass-selected galaxy clusters}
In generating the mass-selected galaxy cluster sample, we start by selecting those haloes with $M_{200} \ge 10^{14} \, M_\odot$ within the $z = 0$ group catalog and then track their central subhalo back from $z = 0$ to $z = 2$ or $1$, using the Main Progenitor Branch of the TNG300 SubLink merger trees via the API. For these (sub)haloes, we store $M_{200}$ at $z = 0$, as well as the group ID and position of the massive progenitor halo at $z = 1, 2$.

We check whether the 3 haloes that are already more massive than $10^{14} \, M_\odot$ at $z = 2$ and 50 haloes that meet the same condition at $z = 1$ are included by this approach. Oddly, a portion are not -- 8 of the massive haloes at $z = 1$ are not returned by this check for massive haloes at $z = 0$. This seems to be an effect of using central membership to trace haloes at different snapshots. These eight massive haloes all evolve so that their centrals are part of a massive halo at $z = 0$, but are not part of the main progenitor branch of the massive halo's central subhalo at $z = 0$.

Before any permutations of the simulation volume, the mass-selected galaxy cluster sample contains 276 haloes at $z = 2$ and 277 haloes at $z = 1$, down from 280 haloes with $M_{200} \ge 10^{14} \, M_\odot$ at $z = 0$.

\subsection{Radius-selected galaxy clusters}

Selecting clusters by their radius at $z = 1$ or $2$ is somewhat more straightforward. We define radius-selected galaxy clusters at each redshift as haloes with $R_{200} \ge 0.75\,h^{-1}$ \textbf{c}Mpc. (See also \citealt{2009MNRAS.394.1469D} for a more in-depth discussion of proto-cluster and cluster structure.)

There are 26 haloes at $z = 2$ and 160 haloes at $z = 1$ that match this criterion.

\subsection{Overlap between cluster catalogs}

Our definitions of mass- and radius-selected galaxy clusters are not mutually exclusive. At $z = 1$, 128 haloes are included in both samples. This dually defined population drops to 24 haloes at $z = 2$.

The overlapping designations for these haloes do not impact our analyses, but we note that this may be of interest in examining the discrepancy between the observed  \ion{H}{i}  flux of mass- and radius-selected clusters or to those looking to assess the validity of one cluster definition over another.

\bsp
\label{lastpage}
\end{document}